\documentstyle[12pt]{article}
\renewcommand{\theequation}{\arabic{section}.\arabic{equation}}
\newcommand{\eqreset}{\setcounter{equation}{0}}

\newtheorem{theorem}{Theorem}
\newtheorem{example}{Example}
\newtheorem{definition}{Definition}
\newtheorem{corollary}{Corollary}

\begin{document}

\title{{\bf Determining equations for higher-order decompositions of exponential operators}
\footnote{The present paper is 
mainly based on the master thesis$^{10}$.}}
\author{Zengo Tsuboi 
%\thanks{{\it 
%Present address : Institute of Physics, College of Arts and Sciences, University of Tokyo, 
%Komaba 3-8-1, Meguro-ku, Tokyo 153, Japan. 
%The present paper is 
%mainly based on the master thesis. }} 
\hspace{0.6cm} and \hspace{0.6cm} Masuo Suzuki \\
{\it Department of Physics, University of Tokyo,} \\
{\it  Bunkyo-ku, Tokyo 113, Japan}}
\date{ }
\maketitle

%\begin{large}

\begin{abstract}
The general decomposition theory of exponential operators is briefly reviewed.
A general scheme to construct independent determining equations for the relevant decomposition parameters is proposed using Lyndon words. Explicit formulas of
the coefficients are derived.
\end{abstract}
\noindent  
Journal-ref: Int. J. Mod. Phys. B9 (1995) 3241-3268 \\
DOI: 10.1142/S0217979295001269
\eqreset
\section{Introduction}

An exponential operator $e^{x(A+B)}$ composed of non-commutable operators
 $A$ and $B$ plays an important role in many fields. Since the sum $A+B$ is 
difficult to diagonalize, we often$^{1-11}$ approximate the exponential 
 operator $e^{x(A+B)}$ as the product of each of the operators $A$ and $B$ : 
\begin{equation}
e^{x(A+B)} = e^{t_{1}A}e^{t_{2}B}e^{t_{3}A}e^{t_{4}B}
 \cdots e^{t_{M}A}+ O(x^{m+1}) .
\end{equation}
This decomposition formula conserves the symmetry, such as unitarity and 
symplecticity. Thus, many people have been using this formula.

\eqreset
\section{Generalized Trotter formulas}

One of the simplest decomposition formulas of the exponential operator
$e^{x(A+B)}$ may be the following Trotter formula$^{11-17}$
\begin{equation}
e^{x(A+B)} = \lim_{n \to \infty} (e^{{\frac{x}{n} }A}e^{{\frac{x}{n} }B})^n .
\end{equation}
The correction of the product in (2.1) for large $n$ is O$(\frac{1}{n})$. That 
is, we have
\begin{equation}
e^{x(A+B)} = (e^{{\frac{x}{n} }A}e^{{\frac{x}{n} }B})^n +  {\rm O}(\frac{1}{n}).
\end{equation}

More generally, if $F_{m}(x)$ is an $m$ th order approximant of $e^{x(A+B)}$,
i.e.
\begin{equation}
e^{x(A+B)} = F_{m}(x) +  {\rm O}(x^{m+1}),
\end{equation}
then we have the following generalized Trotter formula$^{15-17}$ 
\begin{equation}
e^{x(A+B)} = [F_{m}(\frac{x}{n})]^n +  O(\frac{x^{m+1}}{n^m})
\end{equation}
for  $|\frac{x}{n}|<<1$ .

\eqreset
\section{Generalized Baker-Campell-Hausdorff formulas}

The product formula of the two
exponential operators $e^A$ and $e^B$ of the form
\begin{equation}
e^{A}e^{B} = e^{A + B + \frac{1}{2} [A, B] + \cdots } 
\end{equation}
has been well known as the BCH formula$^{18}$.
Here the logarithm of the right hand side of (3.1) is a linear combination of 
basic elements of the free Lie algebra$^{18}$ which is the vector space spanned by the whole set of commutators of $\{ A, B \}$.

Now we consider the following exponential product 
formula$^{1-9}$ :
\begin{equation}
F_{m}(x) = e^{t_{11}A_{1}}e^{t_{12}A_{2}} \cdots e^{t_{1q}A_{q}}
e^{t_{21}A_{1}}e^{t_{22}A_{2}} \cdots e^{t_{2q}A_{q}} 
\cdots \cdots e^{t_{Mq}A_{q}} .
\end{equation} 
Using Friedrichs' theorem$^{18}$, we can extend the Baker-Hausdorff 
theorem in the following$^{19}$.
\begin{theorem}
. The exponential product formula {\rm (3.2)} is expressed in the form
\begin{equation}
F_{m}(x) = \exp (xR_{1} + x^{2}R_{2} + \cdots + x^{n}R_{n} + \cdots), 
\end{equation}
where $R_{n}$ is given by a linear combination of the $n$-th order 
commutators of $\{ A_{j} \}$.
\end{theorem}
 
By the way, it should be remarked that the number of basis elements of degree
$n$ for the operators $\{ A_{j} \}$ ($j = 1, 2, \cdots, r$)
is given by Witt's formula$^{18}$.
\begin{theorem}.
The number of basis elements of degree $n$, $M_{r}(n)$, is expressed by
\begin{equation}
M_{r}(n) = \frac{1}{n}\sum_{d|n} \mu (d)r^{\frac{n}{d}} ,
\end{equation}
where the symbol $d|n$ denotes all common divisors of $n$.
$\mu (d)$ is the M\"obius function : 
$\mu (d)$ is defined for all positive
integers by $\mu (1) = 1$, $\mu (p) = -1$ if $p$ is a prime number,
 $\mu (p^{k}) = 0$ for $k > 1$, and $\mu (cd) = \mu (c)\mu (d)$ if $c, d$
 are coprime integers. 
\end{theorem}

Thus, the condition that $F_{m}(x)$ is an $m$-th order approximant of 
$e^{x(A_{1} + \cdots + A_{q})}$ (i. e., $R_{1} = {\cal H}, R_{2} = \cdots 
R_{m} = 0$) yields $(1 + \sum_{j=2}^{m} M_{q}(j))$ determining 
equations$^{2}$.

Furthermore, we have the following theorem$^{18}$.
\begin{theorem}[Witt's second formula]
The number of independent commutators with each indeterminate
$A_{j}$ containing $n_{j}$ times, $M(n_{1},n_{2}, \cdots, n_{r})$, is given by
the formula.
\begin{equation}
M(n_{1},n_{2}, \cdots, n_{r}) = \frac{1}{\sum_{i=1}^{r} n_{i}} 
\sum_{d|\{ n_{j} \}} \mu (d)
 \frac{(\frac{\sum_{l=1}^{r} n_{l}}{d})!}
{\prod_{k=1}^{r} (\frac{n_{k}}{d})!} .
 \end{equation}
\end{theorem}

\eqreset
\section{General theory of higher-order decomposition of exponential
operators}

Now we consider the following general scheme$^{1-9}$ for constructing
the $m$-th order approximant $F_{m}(x)$ as a product of the $s$-th 
order approximant $Q_{s}^{(j)}(x)$ :
\begin{equation}
F_{m}(x) = Q_{s}^{(1)}(p_{1}x)Q_{s}^{(2)}(p_{2}x) \cdots Q_{s}^{(r)}(p_{r}x),
\end{equation}
where the parameters $\{ p_{j} \}$ satisfy the condition  
\begin{equation}
p_{1} + p_{2} + \cdots + p_{r} = 1 .
\end{equation}
According to Theorem 1, $Q_{s}^{(j)}(x)$ is written as 
\begin{equation}
Q_{s}^{(j)}(x) = \exp (x{\cal H} + x^2R_{j2} + x^3R_{j3} + \cdots ) ,
\end{equation}
where each $R_{jm}$ is a linear combination of the
$m$-th order commutators of $\{ A_{k} \}$.
We have the requirement that $R_{j2} = R_{j3} = \cdots = R_{js} = 0$, and
$R_{j(s+1)} \neq 0$ from the 
condition that $Q_{s}^{(j)}(x)$ is an $s$-th order approximant of 
$e^{x{\cal H}}$.

Moreover, if we assume the following condition of symmetry
\begin{equation} 
Q_{s}^{(j)}(x) Q_{s}^{(j)}(-x) = 1 ,
\end{equation}
then we obtain$^{16}$
\begin{equation}
Q_{s}^{(j)}(x) = \exp (x{\cal H} + x^3R_{j3} + x^5R_{j5} + \cdots )
\end{equation}
In this case, the odd-order approximant $Q_{2k-1}^{(j)}(x)$ is reduced
 to the even-order
one $Q_{2k}^{(j)}(x)$, ($s=2k-1$), namely 
\begin{equation}
Q_{2k-1}^{(j)}(x) = Q_{2k}^{(j)}(x) .
\end{equation}

Now we substitute (4.3) into (4.1), and consequently we obtain$^{2,6,9}$
\begin{eqnarray}
F_{m}(x) &=& \prod_{j=1}^{r} \exp (xp_{j}{\cal H} + (xp_{j})^{2}R_{j2} + 
(xp_{j})^{3}R_{j3} + \cdots ) \\
&=& e^{xH} + \sum_{\{ n_{j} \}} ^{\qquad \prime } \frac{x^{n_{1} + 
2n_{2} + \cdots}}{n_{1}!n_{2}!\cdots}
 {\rm PS}(Y_{1}^{n_{1}} Y_{2}^{n_{2}} \cdots),
\end{eqnarray}
where
\begin{equation}
Y_{1} = \sum_{j=1}^{r} (p_{j}{\cal H}) 
\qquad {\rm and} \qquad 
Y_{n} = \sum_{j=1}^{r} (p_{j}^{n}R_{jn}) \ .
\end{equation}
Here, the symbol $\sum ^{\prime} $ in (4.8) denotes the summation over 
$\{ n_{j} \}$ excluding $n_{2} = n_{3} = \cdots = 0$, and the symbol P denotes 
the {\it time-ordering}$^{2,3}$ operation with respect to the subscript $j$ . 
The symbol S denotes Kubo's symmetrization operation$^{20}$
with respect to the same subscript $j$ as 
\begin{eqnarray}
{\rm S}(R_{jm}^{p}R_{jn}^{q} \cdots ) = \frac{p!q! \cdots }{(p + q + \cdots )!}
\sum {\rm Permu}(R_{jm}^{p}R_{jn}^{q} \cdots )
\end{eqnarray}
for any positive integers $p$ and $q$ ; the symbol "Permu" denotes permutation of the order of
the operators $ \{ R_{jm} \} $ in all possible ways. Here $p_{j}$ and 
$p_{j}^n$ should not be separated before the operations P and S are performed.

Here we give some typical examples of higher-order decomposition
of exponential operators. 

{\bf (1) Nonsymmetric complex decomposition}$^{5}$ \\
   If we choose $\{ Q_{s}^{(j)}(x) \}$ as
\begin{equation}
\{ Q_{s}^{(1)}(x) \} = \{ Q_{s}^{(2)}(x) \} = \cdots =\{ Q_{s}^{(j)}(x) \}
=F_{1}(x)
\end{equation}
with the first order approximant $F_{1}(x)$, then we obtain
\begin{equation}
F_{m}(x) = F_{1}(p_{1}x)F_{1}(p_{2}x) \cdots F_{1}(p_{r}x) .
\end{equation} 

{\bf (2) Symmetric decomposition }$^{2}$ \\
   If we choose $\{ Q_{s}^{(j)}(x) \}$ as
\begin{equation}
\{ Q_{s}^{1}(x) \} = \{ Q_{s}^{2}(x) \} = \cdots =\{ Q_{s}^{j}(x) \}
=S_{2}(x)
\end{equation}
with the second-order symmetric decomposition $S_{2}(x)$ defined by
\begin{equation}
S_{2}(x) =F_{1}(\frac{x}{2})\tilde{F}_{1}(\frac{x}{2}) ,
\end{equation}
 and with the symmetric parameters $\{ p_{j} \}$ satisfying the relation
$p_{r+1-j} = p_{j}$, then we obtain
\begin{equation}
S_{2m}(x) = S_{2m-1}(x) =  S_{2}(p_{1}x)S_{2}(p_{2}x) \cdots S_{2}(p_{r}x) .
\end{equation} 

{\bf (3) Nonsymmetric tilde decomposition}$^{5}$ \\
   If we choose $\{ Q_{s}^{(j)}(x) \}$ as
\begin{equation}
\{ Q_{s}^{(1)}(x) \} = \{ Q_{s}^{(3)}(x) \} = \cdots = F_{1}(x)
\quad {\rm and } \quad
\{ Q_{s}^{(2)}(x) \} = \{ Q_{s}^{(4)}(x) \} = \cdots = \tilde{F}_{1}(x) ,
\end{equation}
 then we obtain 
\begin{equation}
F_{m}(x) = F_{1}(p_{1}x)\tilde{F}_{1}(p_{2}x)
F_{1}(p_{3}x)\tilde{F}_{1}(p_{4}x) \cdots ,
\end{equation} 
and the operators $\{ R_{jn} \}$ in (4.5) are expressed in the form
\begin{equation}
R_{jn} = (-1)^{(j-1)(n-1)}R_{n} .
\end{equation}
Here, $R_{n}$ is independent.
 This nonsymmetric tilde decomposition is very general, 
because if we put $p_{2} = p_{4} = p_{6} = \cdots = 0$ in
(4.17), this reduces to (4.12); if we put 
$p_{2j-1} =p_{2j}, p_{j} = p_{r-j+1}$ and $S_{2}(x) =
 F_{1}(x)\tilde{F}_{1}(x)$ in (4.17), this reduces to (4.15).
By the way, this scheme had been called {\it nonsymmetric real decomposition},
because there exist real parameters $\{ p_{j} \}$ in (4.17) for some cases.
For example, we have$^{21}$
\begin{equation}
F_{2}^{(R)}(x) = e^{\frac{7}{24}xA}e^{\frac{2}{3}xB}e^{\frac{3}{4}xA}
e^{-\frac{2}{3}xB}e^{-\frac{1}{24}xA}e^{xB}.
\end{equation} 

{\bf (4) General recursive scheme}$^{2, 5}$  \\
 If we choose $\{ Q_{s}^{(j)}(x) \}$ in $(4.1)$ as
\begin{equation}
\{ Q_{s}^{(1)}(x) \} = \{ Q_{s}^{(2)}(x) \} = \cdots =\{ Q_{s}^{(j)}(x) \}
=F_{m-l}(x)
\end{equation}
with the $(m-l)$-th order approximant $F_{m-l}(x)$, then we obtain
\begin{equation}
F_{m}(x) = F_{m-l}(p_{1}x)F_{m-l}(p_{2}x) \cdots F_{m-l}(p_{r}x).
\end{equation} 
The case $l = m-1$ corresponds to (4.12).

{\bf (5) Recursive symmetric decomposition }$^{2}$ \\
   If we choose the $(2m-2l)$-th order symmetric decomposition 
$S_{2m-2l}(x)$
as $\{ Q_{s}^{j}(x) \}$ in (4.1) and with the symmetric parameters $\{ p_{j} \}$ satisfying the relation
$p_{r+1-j} = p_{j}$ ,
 then we obtain
\begin{equation}
S_{2m}(x) = S_{2m-1}(x) =  S_{2m-2l}(p_{1}x)S_{2m-2l}(p_{2}x) \cdots 
S_{2m-2l}(p_{r}x) .
\end{equation} 
The case $l = m-1$ corresponds to (4.15).

{\bf (6) Recursive tilde decomposition}$^{5}$ \\
 If we choose $\{ Q_{s}^{(j)}(x) \}$ as
\begin{equation}
\{ Q_{s}^{(1)}(x) \} = \{ Q_{s}^{(3)}(x) \} = \cdots 
=F_{m-l}(x)
\ {\rm and} \ 
\{ Q_{s}^{(2)}(x) \} = \{ Q_{s}^{(4)}(x) \} = \cdots 
=\tilde{F}_{m-l}(x) ,
\end{equation}
 then we obtain
\begin{eqnarray}
F_{m}(x) = F_{m-l}(p_{1}x)\tilde{F}_{m-l}(p_{2}x)F_{m-l}(p_{3}x)
\tilde{F}_{m-l}(p_{4}x) \cdots . 
\end{eqnarray}
The case $l = m-1$ corresponds to (4.17). 

As is easily seen from the above schemes, the following conditions are 
equivalent$^{2,5}$ to each other : 

{\bf (1)} $F_{m}(x)$ is an $m$-th order approximant of $e^{x{\cal H}}$.

{\bf (2)} PS$(Y_{1}^{n_{1}}Y_{2}^{n_{2}} \cdots ) = 0$ for all
$\{ n_{j} \} \in X_{m} $.

{\bf (3)} $g(i_{1},i_{2}, \cdots ) = 0$ for all $(i_{1},i_{2}, \cdots )$ :
the set of number $(i_{1},i_{2}, \cdots )$ is takes an arbitrary permutation of
 \[ (\underbrace{1, 1, \cdots,1}_{n_{1}},\underbrace{2, 2, \cdots,2}_{n_{2}},
 3, \cdots ). \]
Here $ \{ n_{j} \} \in X_{m}$ for a given positive integer set $X_{m}$, 
and $g(i_{1},i_{2}, \cdots )$ is the coefficient of
 $\frac{1}{n_{1}!n_{2}!\cdots}R_{i_{1}}R_{i_{2}}\cdots$ 
in PS$(Y_{1}^{n_{1}}Y_{2}^{n_{2}} \cdots )$, where 
$X_{m}$ is given, for example, as follows :
\\
{\bf (1) Nonsymmetric decomposition} ( see (4.12) amd (4.17) )
\begin{eqnarray}
X_{m} &=& \{ \{ n_{j} \} \in {\bf N}^{\Lambda}|n_{1} + 2n_{2} + \cdots +
mn_{m} \le m , \ {\rm excluding} \nonumber \\ 
& & n_{2} =n_{3} = \cdots = 0 \}\ ; \ 
\Lambda = \{1, 2, \cdots, m \}.
\end{eqnarray}
{\bf (2) Symmetric decomposition} ( see (4.15) )
\begin{eqnarray}
\lefteqn{X_{2m} = X_{2m-1} = \{ \{ n_{j} \} \in {\bf N}^{\Lambda}|n_{1} + 3n_{3}
+\cdots +(2m-1)n_{2m-1}}\nonumber \\
& & =3, 5, \cdots 2m-1, 
 \ {\rm excluding} \ 
 n_{3} =n_{5} = \cdots = 0 \} ;\nonumber \\
& & \Lambda = \{1, 3, 5, \cdots, 2m-1 \}. 
\end{eqnarray}
{\bf (3) Recursive scheme} ( see (4.21) and (4.24) )
\begin{eqnarray}
\lefteqn{X_{m} = \{ \{ n_{j} \} \in {\bf N}^{\Lambda} | n_{1} + (m-l+1)n_{m-l+1} + (m-l+2)n_{m-l+2} }  \nonumber \\
& & = \cdots +mn_{m} \le m , \ 
{\rm excluding} \ n_{m-l+1} =n_{m-l+2} 
= \cdots = 0 \} ;\nonumber \\
& & \Lambda = \{1, m-l+1, m-l+2, \cdots, m \}.
\end{eqnarray}
{\bf (4) Recursive symmetric decomposition} ( see (4.22) )
\begin{eqnarray}
\lefteqn{X_{2m} = X_{2m-1} = 
\{ \{ n_{j} \} \in {\bf N}^{\Lambda} | n_{1} 
+ (2m-2l+1)n_{2m-2l+1} } \nonumber \\
& & + (2m-2l+3)n_{2m-2l+3} + \cdots + (2m-1)n_{2m-1}  
= 2m-2l+1, \nonumber \\ 
& & 2m-2l+3, \cdots 2m-1 , \ 
{\rm excluding} \ n_{2m-2l+1} =n_{2m-2l+3} = \cdots = 0 \} ;\nonumber \\
& & \Lambda = \{1, 2m-2l+1, 2m-2l+3, \cdots, 2m-1 \}.
\end{eqnarray}
As there are a lot of redundant conditions in {\bf (3)}, 
we try to select the minimal independent conditions. 
One of the main purposes in the present paper is to show that the minimal 
independent conditions are given as the coefficients of the 
Lyndon words$^{24}$ generated by the correction terms $\{ R_{j} \}$. 
The details of this statement will be explained later.

Furthermore, there are some general convergence theorems$^{8,22,23}$
in the limit $m \to \infty$ on (4.1).

\eqreset
\section{Lyndon words and free Lie algebras}

Let $X$ be a totally ordered set, $W(X)$ be the set of all words over
the set $X$ and the total order on $X$ is extended lexicographically to $W(X)$.
The Lyndon words $Ly(X)$ over the set $X$ is defined by the set of such elements of
$W(X)$ as have the following properties$^{24}$ :
\begin{quote}
{\bf (1)} A primitive word ( i.e., a word that cannot be expressed as the 
power of another word ). \\
{\bf (2)} A word which is minimal in
 its conjugate class\footnote{The words $x$ and $y$ are conjugate if
$\exists a, b \in W(X)$ such that $x = ab$ and $y = ba $.} 
( lexicographically minimal in its cyclic permutations )
\end{quote}
\begin{example}.
{\rm For} $X = \{ x, y, z \}$ {\rm and} $x < y < z$,
{\rm some of the first few Lyndon words are given as follows} {\rm :}
\begin{eqnarray}
Ly(X) = \{ x, y, z, xy,xz,yz, xxy, xxz, xyy, xyz, xzy, xzz, yyz, yzz, xxxy, 
\cdots \} .
\end{eqnarray}
{\rm The following examples are not Lyndon words} {\rm :}
\begin{eqnarray}
xyxy \quad & & {\rm ( which \ contradicts \ (1) ,\ because \ it \ is 
\ expressed \ as \ a \ power \ of } \ xy ) \qquad \\
 xxyx \quad & & {\rm ( which \ contradicts \ (2) ; } \ xxxy < xxyx ) .
\end{eqnarray}
\end{example}

There are several ways to decompose a Lyndon word as a product of
two Lyndon words$^{24}$ :
\begin{equation}
xxyy = (xxy)(y) = (x)(xyy) .
\end{equation}
We define here the so-called standard factorization$^{24}$ .
 \begin{definition}.
The pair $(l, m)$ for $l, m, \in Ly(X)$ is called 
standard factorization of $w \in Ly(X)$, if $m$ has maximal
length. This factorization is denoted as $\sigma (w)$ .
\end{definition}
\begin{example}.
{\rm For} $X = \{ x, y \}$ {\rm and} $x < y$, {\rm we have}
\begin{eqnarray}
\sigma (xxxyy) = (x,xxyy), \qquad \sigma (xyxyy) = (xy,xyy)
\end{eqnarray}
{\rm The following examples are not standard factorization}  {\rm :}
\begin{eqnarray}
(xxxy,y) ; xxyy \ {\rm \ is \ longer \ than } \ y, \qquad 
(x,yxyy) ; yxyy \notin Ly(X) . 
\end{eqnarray}
\end{example}
There is a bijection$^{24}$ $\lambda$ from the set of Lyndon words $Ly(X)$ to
the set of bases of the free Lie algebra L$(X)$ over the set $X$
as a $K$-free module\footnote{$K$ is a commutative ring with unit element. } 
( namely an additive group on $K$ without any restriction ). 
\begin{definition}.
The bijection $\lambda $ is defined by $\lambda (x)$ satisfying the properties
 that $\lambda (x) = x$ for $x \in X$ and that
\begin{equation}
\lambda (l) = [ \lambda (m), \lambda (n) ]
\end{equation}
for $l \in Ly(X)=X$, where $\sigma (l) = (m,n)$ is the standard factorization
of $l$.
\end{definition}
\begin{example}.
{\rm For} $X = \{ x, y \}$ {\rm and} $x < y$ , {\rm the bijection}
 $\lambda (xxyxyy)$ {\rm of the Lyndon word} $xxyxyy$ {\rm is given by the 
following commutator :}
\begin{eqnarray}
\lambda (xxyxyy) &=& [ \lambda (x), \lambda (xyxyy) ] \nonumber \\
&=& [ x, [ \lambda (xy), \lambda (xyy) ]] \nonumber \\
&=& [ x, [[ \lambda (x), \lambda (y) ], [ \lambda (xy), \lambda (y) ]]] 
\nonumber \\
&=& [ x, [[ x, y ], [[ \lambda(x), \lambda(y) ], y ]]] \nonumber \\
&=& [ x, [[ x, y ], [[ x, y ], y ]]] 
\end{eqnarray}
\end{example}
Now we remark the following theorem$^{24}$.
\begin{theorem}.
The K-module L(X) is spanned by the set $\lambda ( Ly(x) )$.
\end{theorem}
This theorem is very important in our arguments.
\begin{example}.
{\rm For} $X = \{ x, y, z \}$ {\rm and} $x < y < z$, {\rm we have} 
\begin{eqnarray}
[[ x, y ], z ] &=& [ x, [ y, z ]] - [ y, [ x, z ]] \nonumber \\
&=& [ x, [ y, z ]] + [[ x, z ], y ] \nonumber \\
&=& \lambda (xyz) + \lambda (xzy) ,
\end{eqnarray}
{\rm where} $xyz, xzy \in Ly(X) ; \sigma (xyz) = (x,yz)$ {\rm and} $\sigma (xzy) = (xz,y) $.
\end{example}
Next we discuss the set $Ly(X) \cap X^{k}$. Here, $X^k$ denotes a $k$ th order
set of products of elements. Now we write as $Ly^{(k)}(X)=Ly(X) \cap X^k$.
On this set, the following theorem holds$^{24}$.  
\begin{theorem}
For each $l \in Ly^{(k)}(X)=Ly(X) \cap X^{k}, k \ge 1$, we have $\lambda (l) 
= l + r$, 
where $r$ belongs to the submodule\footnote{The set $KW(X)$ is a $K$-module 
generated by $W(X)$, namely an additive group with coefficients of the 
commutative ring $K$ on the set of all worlds over the set $X$.} of $KW(X)$ 
generated by those words
$w \in X^{k}$ such that $l < w$ {\rm :} $r = \sum_{i} k_{i}w_{i}$, $k_{i} \in
K$, $w_{i} \in X^{k}$, $l < w_{i}$.
\end{theorem}
\begin{example}.
{\rm For} $X = \{ x, y \}, x < y, l = xxyy \in Ly^{(4)}(X)=Ly(X) \cap X^{4} $, {\rm we have}\begin{eqnarray}
\lambda ( xxyy ) &=& [ x, [[ x, y ], y ]] \nonumber \\
&=& xxyy - 2xyxy + 2yxyx - yyxx .
\end{eqnarray}
{\rm We find that} $xxyy < xyxy <yxyx < yyxx$ , {\rm and that}
$xxyy, xyxy \in Ly(X)$ {\rm and} $yxyx, yyxx \in X^4-Ly^{(4)}(X)$.
\end{example}
Now we can prove the following theorem.
\begin{theorem}.
Any element in $Ly^{(k)}(X)=Ly(X) \cap X^{k}$ can be written, by theorem 4, as
$\sum_{i=1}^{dim L_{n}(X)} \alpha _{i}\lambda (l_{i}) $,
where $\alpha _{i} \in K$ and $l_{i} \in Ly^{(n)}(X) := Ly(X)$
for $l_{1} < l_{2} < \cdots < l_{dim L_{n}}$ .
Here, $L_{n}(X)$ is an $n$-th order submodule of $L(X)$.
This can be expanded as follows {\rm :}
\begin{eqnarray}
\sum_{i=1}^{dim L_{n}(X) } \alpha _{i}\lambda (l_{i}) = 
\sum_{i=1}^{dim L_{n}(X)} \beta _{i}l_{i} + 
\sum_{j} \gamma _{j}l_{j}^{\prime} ; \quad
l_{j}^{\prime} \in X^{n} - Ly^{(n)}(X) .
\end{eqnarray}
$\beta _{i}$ and $\gamma _{j}$ are first-order homogeneous
polynomials of $\{ \alpha _{i} \}$. Then, we obtain the following statement :
\[ \{ \alpha _{i} \} = \{0\} \Leftrightarrow \{ \beta _{j} \} = \{0\}. \]
\end{theorem}
{\it Proof}.\\
a) The proof of the statement ( $\Rightarrow$ ) is obvious. \\
b) The proof for the statement ( $\Leftarrow$ ) is given as follows.
Let us assume that $\{ \beta _{j} \}$ = $\{0\}$. We have by theorem 5 
\begin{equation}
\lambda ( l_{i} ) = l_{i} + \sum_{j} k_{ij}w_{j}
\end{equation}
where i = 1, 2, $\cdots$, dim$L_{n}(X)$. For any $j$ such that $k_{ij} \ne 0$
for $k_{ij} \in K$, we have  $l_{i} < w_{j}$ for $w_{j} \in X^{n}$. Then we 
obtain
\begin{eqnarray}
\sum_{i=1}^{dim L_{n}(X)} \alpha _{i}l_{i} +
\sum_{i=1}^{dim L_{n}(X)} \alpha _{i}\sum_{j} k_{ij}w_{j} = 
 \sum_{j} \gamma _{j}l_{j}^{\prime} ,
\end{eqnarray}
from (5.11), using $\{ \beta _{j} \}=\{ 0 \}$. 
Since $l_{1} < l_{2} < \cdots < l_{dimL_{n}(X)}$,
we have $l_{1} < w_{j}$ for any $j$. Clearly, only the first term 
on the left hand side contains the element $l_{1}$, and consequently we obtain
 $\alpha _{1}= 0$. Then we assume that 
$\alpha _{1} = \alpha _{2} = \cdots = \alpha _{p-1}$, which means
\begin{eqnarray}
\sum_{i=p}^{dim L_{n}(X)} \alpha _{i}l_{i} +
\sum_{i=p}^{dim L_{n}(X)} \alpha _{i}\sum_{j} k_{ij}w_{j} =  
 \sum_{j} \gamma _{j}l_{j}^{\prime} .
\end{eqnarray}
For any $j$, such that $k_{ij} \ne 0$, $i \ge p$, we have
$l_{p} < w_{j}$. So only the first term on the left hand side
contains the element $l_{p}$, and consequently we obtain $\alpha _{p}$ =$0$.
Therefore we arrive finally at $\{ \alpha _{i} \}$ =$\{ 0 \}$ by mathematical
induction. \\
From Theorem 6, we arrive at the important statement that the relations
 $\{ \gamma _{j} \} = \{ 0 \}$ result from the 
requirement $\{ \alpha _{i} \} = \{0\}$ or $\{ \beta _{j} \} = \{0\}$.
\begin{example}.
{\rm For} $X = \{x, y \}$ {\rm and} $x < y$, {\rm we have}
\begin{eqnarray}
\lefteqn{\alpha _{1} \lambda ( xxyyy ) + \alpha _{2} \lambda ( xyxyy ) 
= \alpha _{1} [ x, [[[ x, y ], y ], y ]] +
 \alpha _{2} [[ x, y ], [[ x, y ], y ]]} \nonumber \\
& &= \alpha _{1}xxyyy + ( -3\alpha _{1} + \alpha _{2} )xyxyy + 
( 3\alpha _{1} - 3\alpha _{2} )xyyxy + 
( -2\alpha_{1} +2\alpha_{2} ) xyyyx \nonumber \\
& & - \alpha _{2}yxxyy + 4\alpha_{2}yxyxy + 
 ( 3\alpha_{1} - 3\alpha_{2} )yxyyx -
 \alpha_{2}yyxxy + \nonumber \\
& & ( -3\alpha_{1} + \alpha_{2} )yyxyx + 
\alpha_{1}yyyxx 
=:\sum_{i=1}^{2} \beta _{i}l_{i} + \sum_{j=1}^{8} \gamma _{j}l_{j}^{\prime}  
\end{eqnarray}
{\rm for} $l_{1} < l_{2}$ {\rm and} 
$l_{1}^{\prime} < l_{2}^{\prime} < \cdots < l_{8}^{\prime}$ .

{\rm If we compare both sides of (5.15), we obtain} 
\begin{equation}
\pmatrix{ \beta _{1} \cr \beta _{2}\ } =
 \pmatrix{ 1 & 0 \cr -3 & 1 }\pmatrix{ \alpha _{1} \cr \alpha _{2} }, \qquad 
\pmatrix{ \alpha _{1} \cr \alpha _{2}\ } =
 \pmatrix{ 1 & 0 \cr 3 & 1 }\pmatrix{ \beta _{1} \cr \beta _{2} },
\end{equation}
{\rm and}
\begin{eqnarray}
\gamma _{1} &=& -3( 2\beta _{1} + \beta _{2} ),
\gamma _{2} = 2( 2\beta _{1} + \beta _{2} ),
\gamma _{3} = -( 3\beta _{1} + \beta _{2} ),
\gamma _{4} = 4( 3\beta _{1} + \beta _{2} ),
\gamma _{5} = \gamma _{1},  \nonumber \\
\gamma _{6} &=& \gamma _{3},
 \gamma _{7} = \beta _{2}, \gamma _{8} = \beta _{1}.
\end{eqnarray}
{\rm Note that} $\{ \alpha _{j} \}$ = $\{ 0 \}$ $\Leftrightarrow$ 
$\{ \beta _{i} \}$ = $\{ 0 \}$, {\rm and the relations} 
$\{ \gamma _{j} \}$ = $\{ 0 \}$
{\rm follow from requirement} $\{ \beta _{i} \}$ = $\{ 0 \}$.
\end{example}

\eqreset
\section{General scheme to construct independent determining equations} 

As briefly mentioned in section 4, there are many redundant conditions
in (3) of section 4  and thus we try here to select the minimal independent
conditions. For this purpose, we note the following theorem.
\begin{theorem}[M. Suzuki$^{2}$ ] 
{\rm PS}$( Y_{1}^{n_{1}} Y_{2}^{n_{2}} \cdots Y_{m}^{n_{m}} )$ is expressed
as a linear combination of basic Lie elements of degree $n$
 {\rm (} $ = n_{1} + n_{2} + \cdots + n_{m}$ {\rm )}
under the condition that 
\begin{equation}
{\rm PS}( Y_{1}^{n_{1}} Y_{2}^{n_{2}} \cdots ) = 0
\end{equation}
for every $\{ n_{j} \} \in X_{m-1}$,
where $X_{m}$ is given, for example, in {\rm (4.25)}, $\dots $. and {\rm (4.28)}.
\end{theorem}

This theorem is absolutely basic in the present arguments of deriving
determining equations of decomposition parameters in general.
Then we obtain the following proposition from 
Theorems 6 and 7.
\begin{theorem}.
The following  conditions are equivalent to each other {\rm :} \\
{\rm (1)} $F_{m}(x)$ is an $m$-th order approximant of $\exp x{\cal H} $.\\
{\rm (2)} The coefficient $g( i_{1}, i_{2} \cdots )$ of the operator
$R_{i_{1}}R_{i_{2}} \cdots$ in 
$\frac{1}{n_{1}!n_{2}! \cdots } {\rm PS}(Y_{1}^{n_{1}}Y_{2}^{n_{2}} \cdots )$
 vanishes for any set of $( i_{1}, i_{2} \cdots )$ such that 
$R_{i_{1}}R_{i_{2}} \cdots$ becomes a Lyndon word generated
by $\{ R_{i} \}$, where the set of numbers $( i_{1}, i_{2} \cdots )$ 
is given by a permutation of 
$( \underbrace{1, \dots, 1}_{n_{1}}, 
\underbrace{2, \dots,2}_{n_{2}}, 3, \dots \dots )$ for $\{ n_{j} \} \in X_{m}$.
\end{theorem}

 R. I. Mclachlan$^{25}$ first pointed out the relevance
of the Lyndon words to studying higher-order decomposition of exponential
operators, and he used them in proving his theorems ( see (A.5) and (A.6)
in Appendix A ).

By the way, the number of independent determining equations i.e., the minimal 
number of the parameter $\{ p_{j} \}$ for the $m$ th order decomposition
is given by the following formula$^{2}$
\begin{equation}
S_{min}(m) = 1 + \sum_{ \{ n_{j} \} \in X_{m}} M( \{ n_{j} \})
\end{equation}
Furthermore, we can simplify (6.2) using (A.5) and (A.6) in Appendix A.

\begin{example}.
\  \\
{\rm (1) Nonsymmetric decomposition ( see (4.12) and (4.16) )} \\
 {\rm We have} \\
$S_{min}(m)=$ 2, 4, 7, 13, 22, 40, 70, 126, 225, 411, 746, 1376, $\dots$ .\\ 
{\rm for} $m=$ 2, 3, 4, $\dots $, respectively. \\
{\rm (2) Symmetric decomposition (see (4.15) ) (} $S_{min}(2k-1)$ = 
$S_{min}(2k)$ {\rm )} \\
 {\rm We have} \\
$S_{min}(m)=$ 2, 4, 8, 16, 34, 74, 164, 374, $\dots$ . \\
{\rm for} $m$ = 3, 5, 7, $\dots$, 17, $\dots$, {\rm respectively}.
\end{example}

\eqreset
\section{Some examples of minimal independent determining equations
 ( or sufficient conditions )}

From Theorem 8 discussed in the preceding section, we obtain easily the 
minimal independent determining equations (or sufficient conditions )$^{2}$,
namely by requiring that all the coefficients 
$g(i_{1},i_{2},\dots )$ of the operators $R_{i_{1}}R_{i_{2}} \cdots $
corresponding to Lyndon words should vanish.
The main results in the present paper are summarized as follows:
\\
{\bf (1) Nonsymmetric decomposition} ( see (4.12) and (4.17) ) \\
The content of $g( \{ i_{j} \} )$ depends on what decomposition we take.
\\
\underline{1 st order} : $g( 1 ) = 1$.
\\
\underline{2 nd order} : $g( 2 ) = 0$.
\\
\underline{3 rd order} : $g( 3 ) = 0$, $g( 1, 2 ) = 0$.
\\
\underline{4 th order} : $g( 4 ) = 0$, $g( 1, 3 ) = 0$, $g( 1, 1, 2) = 0$.
\\
\underline{5 th order} : $g( 5 ) = 0$, $g( 1, 4 ) = 0$, $g( 2, 3 ) = 0$, 
$g( 1, 1, 3 ) = 0$, $g( 1, 2, 2 ) = 0$, $g( 1, 1, 1, 2 ) = 0$.
\\
\underline{6 th order} : $g( 6 ) = 0$, $g( 1, 5 ) = 0$, $g( 2, 4 ) = 0$,
$g( 1, 1, 4 ) = 0$, $g( 1, 2, 3 ) = 0$, $g( 1, 3, 2 ) = 0$. 
$g( 1, 1, 1, 3 ) = 0$, $g( 1, 1, 2, 2 ) = 0$, $g( 1, 1, 1, 1, 2 ) = 0$.
\\
\underline{7 th order} : $g( 7 )=0$, $g( 1, 6 )=0$, $g( 2, 5 )=0$,
 $g( 3, 4 )=0$,$g( 1, 1, 5 )=0$, \\
 $g( 1, 2, 4 )=0$,
 $g( 1, 4, 2 )=0$, $g( 1, 3, 3)=0$, $g( 2, 2, 3 )=0$,
 $g( 1, 1, 1, 4 )=0$, \\
$g( 1, 1, 2, 3 )=0$, $g( 1, 1, 3, 2 )=0$, $g( 1, 2, 1, 3 )=0$, 
$g( 1, 2, 2, 2 )=0$, $g( 1, 1, 1, 1, 3 )=0$, \\
$g( 1, 1, 1, 2, 2 )=0$, $g( 1, 1, 2, 1, 2 )=0$,
 $g( 1, 1, 1, 1, 1, 2 )=0$.
\\
\underline{8 th order} : $g( 8 )=0$, $g( 1, 7 )=0$, $g( 2, 6 )=0$, $g( 3, 5 )=0$,
$g( 1, 1, 6 )=0$, $g( 1, 2, 5 )=0$, $g( 1, 5, 2 )=0$, 
$g( 1, 3, 4 )=0$, $g( 1, 4, 3 )=0$, $g( 2, 2, 4 )=0$, 
$g( 2, 3, 3 )=0$, $g( 1, 1, 1, 5 )=0$, $g( 1, 1, 2, 4 )=0$,
 $g( 1, 1, 4, 2 )=0$, $g( 1, 2, 1, 4 )=0$, $g( 1, 1, 3, 3 )=0$, 
$g( 1, 2, 2, 3 )=0$, $g( 1, 2, 3, 2 )=0$, $g( 1, 3, 2, 2 )=0$, 
$g( 1, 1, 1, 1, 4 )=0$, \\
$g( 1, 1, 1, 2, 3 )=0$, $g( 1, 1, 1, 3, 2 )=0$, $g( 1, 1, 2, 1, 3 )=0$,
 $g( 1, 1, 3, 1, 2 )=0$, \\
$g( 1, 1, 2, 2, 2 )=0$, $g( 1, 2, 1, 2, 2 )=0$, 
$g( 1, 1, 1, 1, 1, 3 )=0$, $g( 1, 1, 1, 1, 2, 2 )=0$, \\
g( 1, 1, 1, 2, 1, 2 )=0, g( 1, 1, 1, 1, 1, 1, 2 )=0.
\begin{equation}
\  
\end{equation}
\\
{\bf (2) Symmetric decomposition} ( see (4.15) )
\\
\underline{1 st order} : $g( 1 )=1$.
\\
\underline{3 rd order} : $g( 3 )=0$.
\\
\underline{5 th order} : $g( 5 )=0$, $g( 1,1,3 )=0$.
\\
\underline{7 th order} : $g( 7 )=0$, $g( 1, 1, 5 )=0$, $g( 1, 3, 3 )=0$, 
$g( 1, 1, 1, 1, 3 )=0$.
\\
\underline{9 th order} : $g( 9 )=0$, $g( 1, 1, 7 )=0$, $g( 1, 3, 5 )=0$, 
$g( 1, 5, 3 )=0$, \\
$g( 1, 1, 1, 1, 5 )=0$, $g( 1, 1, 1, 3, 3 )=0$,
$g( 1, 1, 3, 1, 3 )=0$, $g( 1, 1, 1, 1, 1, 1, 3 )=0$. 
\\
\underline{11 th order} : $g( 11 )=0$, $g( 1, 1, 9 )=0$, 
 $g( 1, 3, 7 )=0$, $g( 1, 7, 3 )=0$,\\
 $g( 1, 5, 5 )=0$, $g( 3, 3, 5 )=0$, 
$g( 1, 1, 1, 1, 7 )=0$, $g( 1, 1, 1, 3, 5 )=0$,\\
 $g( 1, 1, 1, 5, 3 )=0$, $g( 1, 1, 3, 1, 5 )=0$, 
 $g( 1, 1, 5, 1, 3 )=0$, $g( 1, 1, 3, 3, 3 )=0$,\\
 $g( 1, 3, 1, 3, 3 )=0$, 
$g( 1, 1, 1, 1, 1, 1, 5 )=0$, $g( 1, 1, 1, 1, 1, 3, 3 )=0$,\\
 $g( 1, 1, 1, 1, 3, 1, 3 )=0$, $g( 1, 1, 1, 3, 1, 1, 3 )=0$, 
$g( 1, 1, 1, 1, 1, 1, 1, 1, 3 )=0$.
\\
\underline{13 th order} : $g( 13 )=0$, $g( 1, 1, 11 )=0$, $g( 1, 3, 9 )=0$, 
$g( 1, 9, 3 )=0$, $g( 1, 5, 7 )=0$,\\
 $g( 1, 7, 5 )=0$, 
$g( 3, 3, 7 )=0$, $g( 3, 5, 5 )=0$, $g( 1, 1, 1, 1, 9 )=0$,
 $g( 1, 1, 1, 3, 7 )=0$, \\
 $g( 1, 1, 1, 7, 3 )=0$, $g( 1, 1, 3, 1, 7 )=0$,
 $g( 1, 1, 7, 1, 3 )=0$, 
 $g( 1, 1, 1, 5, 5 )=0$, \\
$g( 1, 1, 5, 1, 5 )=0$,
 $g( 1, 1, 3, 3, 5 )=0$,
$g( 1, 1, 3, 5, 3 )=0$, $g( 1, 1, 5, 3, 3 )=0$,\\
 $g( 1, 3, 1, 3, 5 )=0$, 
$g( 1, 3, 1, 5, 3 )=0$, $g( 1, 3, 3, 1, 5 )=0$,
 $g( 1, 3, 3, 3, 3 )=0$,\\
$g( 1, 1, 1, 1, 1, 1, 7 )=0$, $g( 1, 1, 1, 1, 1, 3, 5 )=0$, 
$g( 1, 1, 1, 1, 1, 5, 3 )=0$, \\
$g( 1, 1, 1, 1, 3, 1, 5 )=0$,
$g( 1, 1, 1, 1, 5, 1, 3 )=0$, $g( 1, 1, 1, 3, 1, 1, 5 )=0$,\\
$g( 1, 1, 1, 5, 1, 1, 3 )=0$, $g( 1, 1, 1, 1, 3, 3, 3 )=0$,
$g( 1, 1, 1, 3, 1, 3, 3 )=0$, \\
$g( 1, 1, 1, 3, 3, 1, 3 )=0$,
$g( 1, 1, 3, 1, 1, 3, 3 )=0$, $g( 1, 1, 3, 1, 3, 1, 3 )=0$,\\
$g( 1, 1, 1, 1, 1, 1, 1, 1, 5 )=0$, $g( 1, 1, 1, 1, 1, 1, 1, 3, 3 )=0$,
$g( 1, 1, 1, 1, 1, 1, 3, 1, 3 )=0$, \\
$g( 1, 1, 1, 1, 1, 3, 1, 1, 3 )=0$,
$g( 1, 1, 1, 1, 3, 1, 1, 1, 3 )=0$,\\
$g( 1, 1, 1, 1, 1, 1, 1, 1, 1, 1, 3 )=0$.
\\
\underline{15 th order}: $g( 15 )=0$, $g( 1, 1, 13 )=0$, $g( 1, 3, 11 )=0$,
$g( 1, 11, 3 ) = 0$, \\
$g( 1, 5, 9 ) = 0$, $g( 1, 9, 5 )=0$, $g( 1, 7, 7 ) = 0$, $g( 3, 3, 9 ) = 0$, 
$g( 3, 5, 7 ) = 0$, \\ 
$g( 3, 7, 5 ) = 0$,\\
$g( 1, 1, 1, 1, 11 ) = 0$, $g( 1, 1, 1, 3, 9 ) = 0$, $g( 1, 1, 1, 9, 3 )=0$,
$g( 1, 1, 3, 1, 9 ) = 0$,\\
 $g( 1, 1, 9, 1, 3 ) = 0$,
$g( 1, 1, 1, 5, 7 ) = 0$, $g( 1, 1, 1, 7, 5 ) = 0$,
$g( 1, 1, 5, 1, 7 ) = 0$,\\
 $g( 1, 1, 7, 1, 5 ) = 0$, 
$g( 1, 1, 3, 3, 7 ) = 0$, $g( 1, 1, 3, 7, 3 ) = 0$, 
$g( 1, 1, 7, 3, 3 ) = 0$,\\
 $g( 1, 3, 1, 3, 7 ) = 0$,
$g( 1, 3, 1, 7, 3 ) = 0$, $g( 1, 3, 3, 1, 7 ) = 0$,
$g( 1, 1, 3, 5, 5 ) = 0$, \\
$g( 1, 1, 5, 3, 5 ) = 0$,
$g( 1, 1, 5, 5, 3 ) = 0$, $g( 1, 3, 1, 5, 5 ) = 0$,
$g( 1, 3, 5, 1, 5 ) = 0$, \\
$g( 1, 5, 1, 5, 3 ) = 0$,
$g( 1, 3, 3, 3, 5 ) = 0$, $g( 1, 3, 3, 5, 3 ) = 0$,
$g( 1, 3, 5, 3, 3 ) = 0$, \\
$g( 1, 5, 3, 3, 3 ) = 0$,
$g( 1, 1, 1, 1, 1, 1, 9 )$ $= 0$,
$g( 1, 1, 1, 1, 1, 3, 7 )$ $= 0$,\\
 $g( 1, 1, 1, 1, 1, 7, 3 )= 0$, $g( 1, 1, 1, 1, 3, 1, 7 )= 0$,
 $g( 1, 1, 1, 1, 7, 1, 3 )= 0$,\\
 $g( 1, 1, 1, 3, 1, 1, 7 )= 0$,
 $g( 1, 1, 1, 7, 1, 1, 3 )= 0$, $g( 1, 1, 1, 1, 1, 5, 5 )= 0$,\\
 $g( 1, 1, 1, 1, 5, 1, 5 )$ $= 0$, $g( 1, 1, 1, 5, 1, 1, 5 )= 0$,
$g( 1, 1, 1, 1, 3, 3, 5 )$ $= 0$,\\
 $g( 1, 1, 1, 1, 3, 5, 3 )= 0$,
$g( 1, 1, 1, 1, 5, 3, 3 )$ $= 0$, $g( 1, 1, 1, 3, 1, 3, 5 )= 0$,\\
$g( 1, 1, 1, 3, 1, 5, 3 )$ $= 0$, $g( 1, 1, 1, 5, 1, 3, 3 )$ $= 0$,
$g( 1, 1, 1, 3, 3, 1, 5 )$ $= 0$, \\
$g( 1, 1, 1, 3, 5, 1, 3 )$ $= 0$,
$g( 1, 1, 1, 5, 3, 1, 3 )$ $= 0$, $g( 1, 1, 3, 1, 1, 3, 5 )$ $= 0$,\\
$g( 1, 1, 3, 1, 1, 5, 3 )$ $= 0$, $g( 1, 1, 3, 3, 1, 1, 5 )$ $= 0$,
$g( 1, 1, 3, 1, 3, 1, 5 )$ $= 0$, \\
$g( 1, 1, 3, 1, 5, 1, 3 )$ $= 0$,
$g( 1, 1, 5, 1, 3, 1, 3 )$ $= 0$, 
$g( 1, 1, 1, 3, 3, 3, 3 )$ $= 0$,\\
 $g( 1, 1, 3, 1, 3, 3, 3 )$ $= 0$,
$g( 1, 1, 3, 3, 1, 3, 3 )$ $= 0$,
 $g( 1, 1, 3, 3, 3, 1, 3 )$ $= 0$,\\
$g( 1, 3, 1, 3, 1, 3, 3 )$ $= 0$,
$g( 1, 1, 1, 1, 1, 1, 1, 1, 7 )$ $= 0$,
$g( 1, 1, 1, 1, 1, 1, 1, 3, 5 )$ $= 0$,\\
 $g( 1, 1, 1, 1, 1, 1, 1, 5, 3 )$ $= 0$,
$g( 1, 1, 1, 1, 1, 1, 3, 1, 5 )$ $= 0$,
 $g( 1, 1, 1, 1, 1, 1, 5, 1, 3 )$ $= 0$,\\
$g( 1, 1, 1, 1, 1, 3, 1, 1, 5 )$ $= 0$,
 $g( 1, 1, 1, 1, 1, 5, 1, 1, 3 )$ $= 0$,
$g( 1, 1, 1, 1, 3, 1, 1, 1, 5 )$ $= 0$,\\
 $g( 1, 1, 1, 1, 5, 1, 1, 1, 3 )$ $= 0$,
$g( 1, 1, 1, 1, 1, 1, 3, 3, 3 )$ $= 0$,
 $g( 1, 1, 1, 1, 1, 3, 1, 3, 3 )$ $= 0$,\\
$g( 1, 1, 1, 1, 1, 3, 3, 1, 3 )$ $= 0$,
 $g( 1, 1, 1, 1, 3, 1, 1, 3, 3 )$ $= 0$,
$g( 1, 1, 1, 1, 3, 1,3, 1, 3 )= 0$,\\
 $g( 1, 1, 1, 1, 3, 3, 1, 1, 3 )= 0$,
$g( 1, 1, 1, 3, 1, 1, 1, 3, 3 )= 0$,
 $g( 1, 1, 1, 3, 1, 1, 3, 1, 3 )= 0$,\\
$g( 1, 1, 1, 3, 1, 3, 1, 1, 3 )= 0$,
$g( 1, 1, 1, 1, 1, 1, 1, 1, 1, 1, 5 )= 0$,\\
$g( 1, 1, 1, 1, 1, 1, 1, 1, 1, 3, 3 )= 0$,
$g( 1, 1, 1, 1, 1, 1, 1, 1, 3, 1, 3 )= 0$,\\
$g( 1, 1, 1, 1, 1, 1, 1, 3, 1, 1, 3 )= 0$,
$g( 1, 1, 1, 1, 1, 1, 3, 1, 1, 1, 3 )= 0$, \\
$g( 1, 1, 1, 1, 1, 3, 1, 1, 1, 1, 3 )= 0$,
$g( 1, 1, 1, 1, 1, 1, 1, 1, 1, 1, 1, 1, 3 )= 0 $.
\begin{equation}
 \ 
\end{equation}
\\
In the above equations (7.1), we have $g(1, 2, 3)=0$ and $g(1, 3, 2)=0$ (and
 we do not need the third equation $g(3, 1, 2)=0$ ), because we have only
two Lyndon words of level 3, $xyz$ and $xzy$ for the three elements $x, y$
 and $z$, namely $M(1, 1, 1)=2$. Similarly note that 
\begin{eqnarray}
 M(n,1) &=& 1 \ {\rm for} \ n \ge 2, \ M(3,1,1)=4, \ M(2,3)=2,
 \nonumber \\
 M(5,2) &=& 3, \ M(2,2,1)=6, \ M(5,1,1)=6,\  M(4,3)=5, \nonumber \\
 M(7,2) &=& 5, \  M(4,2,1)=15,\  M(7,1,1)=8, \nonumber \\
 M(6,3) &=& 9, \  M(9,2)=5, \dots .
 \ 
\end{eqnarray}

\eqreset
\section{Explicit formulas on the coefficients 
of the time-ordered symmetrized operators 
PS$( Y_{1}^{n_{1}}Y_{2}^{n_{2}}$ $\cdots )$ }

Explicit formulas on the coefficients of 
PS$( Y_{1}^{n_{1}}Y_{2}^{n_{2}}$ $\cdots$ $Y_{m}^{n_{m}})$ have been known only
for small $m$. We explain here the derivation of it for arbitrary $m$.
Before we explain the general case, we explain the case $m$ = 3.
\begin{example}.
\begin{eqnarray}
{\rm PS}( Y_{1}Y_{2} ) &=& {\rm PS}( \{\sum_{i} p_{i}R_{1} \}
\{ \sum_{j} p_{j}^{2}R_{2} \} ) \\
&=& {\rm PS} \{ \sum_{i} ( p_{i}R_{1} )( p_{i}^{2}R_{2} ) +
 \sum_{i<j} ( p_{i}R_{1} )( p_{j}^{2}R_{2} ) +
 \sum_{i>j} ( p_{i}R_{1} )( p_{j}^{2}R_{2} ) \} \\
&=& {\rm S} \{ \sum_{i} ( p_{i}R_{1} )( p_{i}^{2}R_{2} ) +
 \sum_{i<j} [ ( p_{i}R_{1} )( p_{j}^{2}R_{2} ) +
 ( p_{i}^{2}R_{2} )( p_{j}R_{1} ) ] \} \\
&=& \sum_{i} \frac{p_{i}^{3}}{2!} ( R_{1}R_{2} + R_{2}R_{1} ) +
 \sum_{i<j} ( p_{i}p_{j}^{2}R_{1}R_{2} +
 p_{i}^{2}p_{j}R_{2}R_{1} ) \\
&=& ( \frac{1}{2} \sum_{i} p_{i}^{3} + \sum_{i<j} p_{i}p_{j}^{2} )R_{1}R_{2} +
( \frac{1}{2} \sum_{i} p_{i}^{3} + \sum_{i<j} p_{i}^{2}p_{j} )R_{2}R_{1}.        \end{eqnarray}
{\rm We must not separate the parameters} $\{ p_{j} \}$
{\rm from the operators} $\{ R_{j} \}$ {\rm before the operations P and S 
are performed. Our explicit procedures in Eqs. (8.2) - (8.5) are the following :}\\
{\rm (8.2) :} to separate terms for $i = j$ from the remaining terms for 
$i \ne j$,
\\
{\rm (8.3) :} to operate {\rm P} to the terms for $i \ne j$ with respect to 
the subscripts $i$ and $j$ ,
\\
{\rm (8.4) :} to operate {\rm S} to the terms for $i=j$,
\\
and
\\
{\rm (8.5) :} to collect the same type of the operators.
\end{example}
If we generalize Example 9, we obtain the following theorem :
\begin{theorem}.
{\rm (} The case $R_{j\beta}$ = $R_{\beta}$ {\rm (} $R_{j\beta}$ is independent of $j$ {\rm ) ) :} 
The coefficient of $R_{i_{1}}R_{i_{2}} \cdots R_{i_{n}}$ in 
${\rm PS}( Y_{1}^{n_{1}}Y_{2}^{n_{2}} \cdots Y_{m}^{n_{m}})$ can be expressed
explicitly as 
\begin{eqnarray}
n_{1}!n_{2}! \cdots n_{m}! \sum_{\alpha}^{n}
 \sum_{\{ t_{j} \} \in Z_{( \alpha, n )}} \frac{1}{t_{1}!t_{2}! \cdots 
t_{\alpha}!} \sum_{k_{1}<k_{2}< \cdots <k_{\alpha}} 
p_{k_{1}}^{i_{1} + i_{2} + \cdots + i_{t_{1}}} \nonumber \\
\times p_{k_{2}}^{i_{t_{1} + 1} + i_{t_{1} + 2} + \cdots + i_{t_{1} + t_{2}}}
\cdots
p_{k_{\alpha}}^{i_{t_{1} + t_{2} + \cdots + t_{\alpha - 1} +1} + 
i_{t_{1} + t_{2} + \cdots + t_{\alpha - 1} + 2} + 
\cdots + i_{n}}
\end{eqnarray}
where $ \ n=n_{1} +n_{2} + \cdots + n_{n}$, 
the set of numbers $( i_{1}, i_{2},\cdots i_{n} )$ 
takes an arbitrary permutation of 
$( \underbrace{1, 1, \dots, 1}_{n_{1}} \underbrace{2,2, \dots, 2}_{n_{2}}
\dots \dots \underbrace{m, m, \dots, m}_{n_{m}} )$ and, \\
$Z_{( \alpha, n )}$ = $\{ \{t_{j} \} | \sum_{j=1}^{\alpha} t_{j} =
n ; 1 \le t_{j} \le n ; t_{j} \in {\bf Z} \}$.
\end{theorem}
Proof.
\begin{eqnarray}
\lefteqn{{\rm PS}( Y_{1}^{n_{1}}Y_{2}^{n_{2}} \cdots Y_{m}^{n_{m}})  
= {\rm PS}\{ ( \sum_{j_{1}} p_{j_{1}}R_{1} )^{n_{1}}
( \sum_{j_{2}} p_{j_{2}}R_{2} )^{n_{2}} \cdots 
( \sum_{j_{1}} p_{j_{m}}R_{m} )^{n_{m}} \}} \nonumber \\
& & = {\rm PS}\{ \sum_{j_{1}^{(1)}} \sum_{j_{1}^{(2)}} \cdots
 \sum_{j_{1}^{(n_{1})}} \sum_{j_{2}^{(1)}} \cdots \sum_{j_{2}^{(n_{2})}}
\cdots\cdots
 \sum_{j_{m}^{(1)}} \cdots \sum_{j_{m}^{(n_{m})}}
\underbrace{ ( p_{j_{1}^{(1)}}R_{1} )( p_{j_{1}^{(2)}}R_{1} )
\cdots ( p_{j_{1}^{(n_{1})}}R_{1} ) }_{n_{1}} \nonumber \\
& & \underbrace{ ( p_{j_{2}^{(1)}}^{2}R_{2} )( p_{j_{2}^{(2)}}^{2}R_{2} )
\cdots ( p_{j_{2}^{(n_{2})}}^{2}R_{2} ) }_{n_{2}} \cdots\cdots 
\underbrace{ ( p_{j_{m}^{(1)}}^{m}R_{m} )( p_{j_{m}^{(2)}}^{m}R_{m} )
\cdots ( p_{j_{m}^{(n_{m})}}^{m}R_{m} ) }_{n_{m}} \}  \\
& & = PS \{ \sum_{\alpha = 1}^{n} \sum_{ \{ n_{i}^{(j)} \} \in 
W( \alpha, \{ n_{j} \} )} \frac{n_{1}!}
{n_{1}^{(1)}!n_{1}^{(2)}! \cdots n_{1}^{(\alpha)}!}
\frac{n_{2}!}{n_{2}^{(1)}!n_{2}^{(2)}! \cdots n_{2}^{(\alpha)}!}
\cdots \frac{n_{m}!}{n_{m}^{(1)}!n_{m}^{(2)}! \cdots n_{m}^{(\alpha)}!} 
\nonumber \\
& & \times \sum_{k_{1} < k_{2} < \cdots < k_{\alpha}} ( p_{k_{1}}R_{1} )^{n_{1}^{(1)}}
( p_{k_{1}}^{2}R_{2} )^{n_{2}^{(1)}} \cdots 
( p_{k_{1}}^{m}R_{m} )^{n_{m}^{(1)}} 
( p_{k_{2}}R_{1} )^{n_{1}^{(2)}}( p_{k_{2}}^{2}R_{2} )^{n_{2}^{(2)}}
\cdots \nonumber \\
& & \times ( p_{k_{2}}^{m}R_{m} )^{n_{m}^{(2)}} 
\cdots \cdots ( p_{k_{\alpha}}R_{1} )^{n_{1}^{(\alpha)}}
( p_{k_{\alpha}}^{2}R_{2} )^{n_{2}^{(\alpha)}} 
 \cdots( p_{k_{\alpha}}^{m}R_{m} )^{n_{m}^{(\alpha)}} 
\end{eqnarray}
where $W( \alpha, \{ n_{j} \} )$ = $\{ \{ n_{i}^{(j)} \} | \sum_{j = 1}^{n}
n_{i}^{(j)} = n_{i}, \  ( i = 1, 2, \dots, m ) ; 1 \le \sum_{i = 1}^{m}
n_{i}^{(j)}, \ ( j = 1, 2, \dots, \alpha ) ; 0 \le n_{i}^{(j)} \le n _{(i)} \}$  
From (8.8), we find such factors as 
$\frac{n_{1}!}{n_{1}^{(1)}!n_{1}^{(2)}!\cdots n_{1}^{(\alpha)}!}$. 
This is the number of combinations such that the number of $\{ p_{j} \}$ 
with the same suffix $j_{1}^{(s)}$ is $k_{s}$ in (8.7). 
On the other hand, a rearrangement of 
$n= n_{1} + n_{2} + \cdots + n_{m}$ suffixes 
$( j_{1}^{(1)}, j_{1}^{(2)}, \dots, j_{1}^{(n_{1})},
j_{2}^{(1)}, j_{2}^{(2)}, \dots, j_{2}^{(n_{2})},
\dots\dots, j_{m}^{(1)}, j_{m}^{(2)}, \dots, j_{m}^{(n_{m})} )$
in (8.8) gives the following set :
\begin{eqnarray}
 ( \underbrace{k_{1}, \dots, k_{1}}_
{n_{1}^{(1)} + n_{2}^{(1)} + \cdots + n_{m}^{(1)}},
\underbrace{k_{2}, \dots, k_{2}}_
{n_{1}^{(2)} + n_{2}^{(2)} + \cdots + n_{m}^{(2)}},
\dots \dots,
\underbrace{k_{\alpha}, \dots, k_{\alpha}}_
{n_{1}^{(\alpha)} + n_{2}^{(\alpha)} + \cdots + n_{m}^{(\alpha)}} )  
\end{eqnarray}
where $k_{1} < k_{2} < \cdots < k_{\alpha}$. After all, the
suffixes $\{ j_{p}^{(q)} \}$ split into $\alpha$ different sectors. 
Therefore, there exist at least one suffix in each sector. Thus we 
need the conditions $1 \le \sum_{i = 1}^{m} n_{i}^{(j)} $
$( j = 1, 2, \dots, \alpha )$.
Because of the operation of S, such factors as 
\begin{eqnarray}
\frac{n_{1}^{(1)}!n_{2}^{(1)}! \cdots n_{m}^{(1)}!}
{(n_{1}^{(1)} + n_{2}^{(1)} + \cdots + n_{m}^{(1)})!} \cdot
\frac{n_{1}^{(2)}!n_{2}^{(2)}! \cdots n_{m}^{(2)}!}
{(n_{1}^{(2)} + n_{2}^{(2)} + \cdots + n_{m}^{(2)})!} \cdots \nonumber \\
\times \frac{n_{1}^{(\alpha)}!n_{2}^{(\alpha)}! \cdots n_{m}^{(\alpha)}!}
{(n_{1}^{(\alpha)} + n_{2}^{(\alpha)} + \cdots + n_{m}^{(\alpha)})!}
\end{eqnarray}
appear in (8.8), when it is rearranged in the above way. Thus, (8.8) is
 reduced to the following formula
\begin{eqnarray}
\lefteqn{n_{1}!n_{2}! \cdots n_{m}!
\sum_{\alpha = 1}^{n} \sum_{ \{ n_{i}^{(j)} \} \in 
W( \alpha, \{ n_{j} \} )} \frac{1}{(n_{1}^{(1)} + n_{2}^{(1)}+ 
 \cdots n_{m}^{(1)})!} } \nonumber \\
& & \times \frac{1}{(n_{1}^{(2)} + n_{2}^{(2)}+ 
 \cdots n_{m}^{(2)})!} \cdots 
\frac{1}{(n_{1}^{(\alpha)} + n_{2}^{(\alpha)}+ \cdots n_{m}^{(\alpha)})!}
\sum_{k_{1} < k_{2} < \cdots < k_{\alpha}} \nonumber \\
& & \times p_{k_{1}}^{n_{1}^{(1)} + 2n_{2}^{(1)} + \cdots + mn_{m}^{(1)}}
 p_{k_{2}}^{n_{1}^{(2)} + 2n_{2}^{(2)} + \cdots + mn_{m}^{(2)}}
\cdots p_{k_{\alpha}}^{n_{1}^{(\alpha)} + 2n_{2}^{(\alpha)} + 
\cdots + mn_{m}^{(\alpha)}} \nonumber \\
& & \times {\rm Permu} ( R_{1}^{n_{1}^{(1)}}R_{2}^{n_{2}^{(1)}} \cdots
R_{m}^{n_{m}^{(1)}} )
{\rm Permu} ( R_{1}^{n_{1}^{(2)}}R_{2}^{n_{2}^{(2)}} \cdots
R_{m}^{n_{m}^{(2)}} ) \cdots \nonumber \\
& & \times \cdots {\rm Permu} ( R_{1}^{n_{1}^{(\alpha)}}R_{2}^{n_{2}^{(\alpha)}}\cdots R_{m}^{n_{m}^{(\alpha)}} ).
\end{eqnarray}
Now we put $\ n_{1}^{(1)} + n_{2}^{(1)} + \cdots + n_{m}^{(1)}$ = $t_{1}$,
$\ n_{1}^{(2)} + n_{2}^{(2)} + \cdots + n_{m}^{(2)}$ = $t_{2}$ ,
$\dots$, and
$\ n_{1}^{(\alpha)} + n_{2}^{(\alpha)} + \cdots + n_{m}^{(\alpha)}$ = 
$t_{\alpha}$. Our desired sum of the coefficients of the operator
\begin{eqnarray}
R_{i_{1}}R_{i_{2}} \cdots R_{i_{n}} &=& R_{i_{1}}R_{i_{2}} \cdots R_{i_{t_{1}}}
R_{i_{t_{1} +1}}R_{i_{t_{1} + 2}} \cdots R_{i_{t_{1} + t_{2}}}
 \cdots \nonumber \\
& & \times \cdots R_{i_{t_{1} + t_{2} + \cdots + t_{\alpha - 1} + 1}}
R_{i_{t_{1} + t_{2} + \cdots + t_{\alpha - 1} + 2}} \cdots R_{i_{n}}
\end{eqnarray}
appearing in Eq. (8.11) 
are obtained by adding only the coefficients of the terms with each 
Permu$( R_{1}^{n_{1}^{(\beta)}}R_{2}^{n_{2}^{(\beta)}}$ $\cdots$
$R_{m}^{n_{m}^{(\beta)}} )$ ( $\beta$ = 1, 2, $\dots$, $\alpha$ )
containing the following type
\begin{equation}
R_{i_{t_{1} + t_{2} + \cdots + t_{\beta - 1} + 1}}
R_{i_{t_{1} + t_{2} + \cdots + t_{\beta - 1} + 2}}
\cdots 
R_{i_{t_{1} + t_{2} + \cdots + t_{\beta - 1} + t_{\beta}}}.
\end{equation}
If we take notice of the fact that 
\begin{eqnarray}
n_{1}^{(\beta)} + 2n_{2}^{(\beta)} + 
\cdots + mn_{m}^{(\beta)} &=& i_{t_{1} + t_{2} + \cdots + t_{\beta - 1} + 1} + i_{t_{1} + t_{2} + \cdots + t_{\beta - 1} + 2} + \cdots \nonumber \\
& & \cdots + i_{t_{1} + t_{2} + \cdots + t_{\beta}} \quad
( \beta = 1, 2, \dots, \alpha ),
\end{eqnarray} 
then we obtain the coefficient (8.6).

We can realize the meaning of $\{ n_{i}^{(j)} \}$
from the following diagram.
\[ \begin{array}{cccccc}
n_{1}^{(1)} & n_{1}^{(2)} & \cdots & n_{1}^{(\alpha)} & 
\stackrel{sum}{\Longrightarrow} & n_{1} \\
n_{2}^{(1)} & n_{2}^{(2)} & \cdots & n_{2}^{(\alpha)} & 
\stackrel{sum}{\Longrightarrow} & n_{2} \\
\vdots      &    \vdots  &        & \vdots      & \vdots & \vdots \\
n_{m}^{(1)} & n_{m}^{(2)} & \cdots & n_{m}^{(\alpha)} & 
\stackrel{sum}{\Longrightarrow} & n_{m} \\
sum\big\Downarrow & sum\big\Downarrow & \cdots  & sum\big\Downarrow & & 
sum\big\Downarrow \\
t_{1} & t_{2} & \cdots & t_{\alpha} & \stackrel{sum}{\Longrightarrow}
 & n
\end{array} \]

\begin{theorem}.
{\rm (} the case $R_{j\beta}$ = $( -1 )^{( j - 1 )( \beta - 1 )}R_{\beta}$ {\rm
 ) (} see $(4.18)$ {\rm ) :}
The coefficient of $R_{i_{1}}R_{i_{2}} \cdots $ $R_{i_{n}}$ in 
PS$( Y_{1}^{n_{1}}Y_{2}^{n_{2}} \cdots Y_{m}^{n_{m}})$ can be expressed 
explicitly as 
\begin{eqnarray}
n_{1}!n_{2}! \cdots n_{m}! \sum_{\alpha}^{n}
 \sum_{\{ t_{j} \} \in Z_{( \alpha, n )}} \frac{1}{t_{1}!t_{2}! \cdots 
t_{\alpha}!} \sum_{k_{1}<k_{2}< \cdots <k_{\alpha}} 
\large\epsilon ( \alpha ; \{ t_{\beta} \} ; \{ k_{\gamma} \} ) \nonumber \\
\times p_{k_{1}}^{i_{1} + i_{2} + \cdots + i_{t_{1}}}
p_{k_{2}}^{i_{t_{1} + 1} + i_{t_{1} + 2} + \cdots + i_{t_{1} + t_{2}}}
\cdots 
\cdots p_{k_{\alpha}}^{i_{t_{1} + t_{2} + \cdots + t_{\alpha - 1} +1} + 
i_{t_{1} + t_{2} + \cdots + t_{\alpha - 1} + 2} + 
\cdots + i_{n}}
\end{eqnarray}
\begin{eqnarray}
\large\epsilon ( \alpha ; \{ t_{\beta} \} ; \{ k_{\gamma} \} ) &=&
( -1 )^{\{ ( k_{1} - 1 )( i_{1} + i_{2} + \cdots + i_{t_{1}} - t_{1} ) + 
( k_{2} - 1 )( i_{t_{1} + 1} + i_{t_{1} + 2} + 
\cdots + i_{t_{1} + t_{2}} - t_{2} ) + \cdots } \nonumber \\
& & ^{+ ( k_{\alpha} - 1 )( i_{t_{1} + t_{2} + \cdots + t_{\alpha - 1} + 1} + 
i_{t_{1} + t_{2} + \cdots + t_{\alpha - 1} + 2} + 
\cdots + i_{t_{n}} - t_{\alpha} ) \}}
\end{eqnarray}
where $n=n_{1} +n_{2} + \cdots + n_{n}$, the set of number
 $( i_{1}, i_{2},\cdots i_{n} )$ 
takes an arbitrary permutation of $( \underbrace{1, 1, \dots, 1}_{n_{1}}
\underbrace{2,2, \dots, 2}_{n_{2}},
\dots\dots, \underbrace{m, m, \dots, m}_{n_{m}} )$ and 
$Z_{( \alpha, n )}$ = $\{ \{t_{j} \} | \sum_{j=1}^{\alpha} t_{j} =
n ; 1 \le t_{j} \le n ; t_{j} \in {\bf Z} \}$.
\end{theorem}
Proof.\\
We can prove this theorem almost in the same way as for Theorem 9 .

\eqreset
\section{Simplification of the coefficients}

We can simplify the coefficient $g( i_{1}, i_{2},$ $\dots$, $i_{n} )$
of $R_{i_{1}}R_{i_{2}}$ $\cdots$ $R_{i_{n}}$ in 
$ \frac{ {\rm PS}( Y_{1}^{n_{1}}Y_{2}^{n_{2}} \cdots Y_{m}^{n_{m}})}
{n_{1}!n_{2}! \cdots n_{m}!} $ 
where $n=n_{1} +n_{2} +$ $\cdots$ $+ n_{n}$ and the set 
$( i_{1}, i_{2},$ $\cdots$, $i_{n} )$ takes an arbitrary permutation of 
( $\underbrace{1, 1, \dots, 1}_{n_{1}} $, 
$\underbrace{2,2, \dots, 2}_{n_{2}} $,
$\dots \dots$ $\underbrace{m, m, \dots, m}_{n_{m}} $ ) ; $R_{1} := {\cal H} $ :
\begin{definition}
\ \\
$(1)$ For n=1,
\begin{equation}
 f( i_{j} ) := g( i_{j} )
\end{equation}
$(2)$ For $n \ge 2 $, the functions $f( i_{1}, i_{2}, \dots, i_{n} )$
are defined recursively as
\begin{eqnarray}
f( i_{1}, i_{2}, \dots, i_{n} ) &:=& g( i_{1}, i_{2}, \dots, i_{n} ) - 
\sum_{\alpha}
 \sum_{\{ t_{j} \} \in Z_{( \alpha, n )}} 
 \frac{1}{t_{1}!t_{2}! \cdots 
t_{\alpha}!} f( i_{1} + i_{2} + \cdots + i_{t_{1}}, 
\nonumber  \\ 
& & i_{t_{1} + 1} + i_{t_{1} + 2} + \cdots + i_{t_{1} + t_{2}},
\dots \dots, i_{t_{1} + t_{2} + \cdots + t_{\alpha - 1} +1} \nonumber \\
& & + i_{t_{1} + t_{2} + \cdots + t_{\alpha - 1} + 2} + \cdots + i_{n} ) ,
\end{eqnarray}
\end{definition}
where
$Z_{( \alpha, n )}$ = $\{ \{t_{j} \} | \sum_{j=1}^{\alpha} t_{j} =
n ; 1 \le t_{j} \le n ; t_{j} \in {\bf Z} \}$ \\
( a ) Nonsymmetric complex decomposition (4.13) : 
\[ \alpha = 1, 2, \dots, n - 1. \]
( b ) Nonsymmetric tilde decomposition (4.17) : \\
$\{ t_{j} \}$ are odd numbers ; $ \alpha = 1, 3, 5, \dots, n - 2$ 
in the case in which $n$ is an odd number and $\alpha = 2, 4, 6, \dots, n - 2$
in the case in which $n$ is an even number. \\
( c ) Symmetric decomposition (4.15) : \\
$\{ t_{j} \}$ are odd numbers\footnote{If we subtract 
$n_{1} + n_{3} + \cdots + n_{2k-1}$ = $n$ from 
$n_{1} + 3n_{3} + \cdots + ( 2k - 1 )n_{2k-1}$ = $2k - 1$
on both sides, we obtain $2( n_{3} + 2n_{5} + \cdots + ( k - 1 )n_{2k-1} )$ = 
$2k - 1 - n$.} and
$\alpha = 1, 3, 5, \dots, n - 2$. 

From the definition 3 , we obtain the following example.

\begin{example}
\ \\
{\rm ( a ) Nonsymmetric complex decomposition :} 
\begin{equation}
n = 1 : \qquad f( i_{1} ) = \sum_{k_{1} < k_{2} < \cdots < k_{n}} 
p_{k_{1}}^{i_{1}}p_{k_{2}}^{i_{2}} \cdots p_{k_{n}}^{i_{n}}.
\end{equation}
{\rm ( b )   Nonsymmetric tilde decomposition :} 
\begin{equation}
n = 1 :\ f( i_{1} ) = ( -1 )^{m - 1}\sum_{k} ( -1 )^{( i_{1} - 1 )k}p_{k}^{i_{1}}. 
\end{equation}
\begin{eqnarray} 
\lefteqn{n = 2 :\qquad f( i_{1},i_{2} ) = ( -1 )^{m}\{ \frac{1}{2}
\sum_{k} ( -1 )^{( i_{1} + i_{2} )k}
p_{k}^{i_{1} + i_{2}} } \nonumber \\
& & + \sum_{k_{1} < k_{2}} 
( -1 )^{( i_{1} - 1 )k_{1} + ( i_{2} - 1)k_{2} )}
p_{k_{1}}^{i_{1}}p_{k_{2}}^{i_{2}} \}. 
\end{eqnarray}
\begin{eqnarray}
\lefteqn{n = 3 : \qquad f( i_{1},i_{2}, i_{3} ) = ( -1 )^{m - 1}\{
\sum_{k_{1} <k_{2} <k_{3} } ( -1 )^{( i_{1} - 1 )k_{1} +
( i_{2} - 1 )k_{2} + ( i_{3} - 1 )k_{3}} 
p_{k_{1}}^{i_{1}}p_{k_{2}}^{i_{2}}p_{k_{3}}^{i_{3}}  \qquad } \nonumber \\
& & + \sum_{k_{1} < k_{2}} \frac{1}{2} 
[ ( -1 )^{( i_{1} - 1 )k_{1} + ( i_{2} + i_{3} )k_{2} )}
p_{k_{1}}^{i_{1}}p_{k_{2}}^{i_{2} + i_{3}} + 
( -1 )^{( i_{1} + i_{2} )k_{1} + i_{3}k_{2} )}
p_{k_{1}}^{i_{1} + i_{2}}p_{k_{2}}^{i_{3}} ] \}.
\end{eqnarray}
{\rm ( c ) Symmetric decomposition} 
\begin{equation}
n =1 : \qquad f( i_{1} ) = \sum_{k}p_{k}^{i_{1}} .
\end{equation}
\begin{equation}
n=3 :\quad f( i_{1},i_{2}, i_{3} ) = \sum_{k_{1} <k_{2} <k_{3} } 
p_{k_{1}}^{i_{1}}p_{k_{2}}^{i_{2}}p_{k_{3}}^{i_{3}} + 
\sum_{k_{1} < k_{2}} \frac{1}{2} (
p_{k_{1}}^{i_{1}}p_{k_{2}}^{i_{2} + i_{3}} +
p_{k_{1}}^{i_{1} + i_{2}}p_{k_{2}}^{i_{3}} ) . 
\end{equation}
\begin{eqnarray}
\lefteqn{n = 5 : \quad f( i_{1},i_{2}, i_{3}, i_{4}, i_{5} ) =
 \sum_{k_{1} <k_{2} <k_{3} <k_{4} <k_{5}} 
p_{k_{1}}^{i_{1}}p_{k_{2}}^{i_{2}}p_{k_{3}}^{i_{3}}p_{k_{4}}^{i_{4}}
p_{k_{5}}^{i_{5}} } \nonumber \\
& & + \sum_{k_{1} < k_{2} < k_{3} < k_{4}} \frac{1}{2} (
p_{k_{1}}^{i_{1}}p_{k_{2}}^{i_{2}}p_{k_{3}}^{i_{3}}
p_{k_{4}}^{i_{4} + i_{5}} +
p_{k_{1}}^{i_{1}}p_{k_{2}}^{i_{2}}
p_{k_{3}}^{i_{3} + i_{4}}p_{k_{5}}^{i_{5}} +
p_{k_{1}}^{i_{1}}p_{k_{2}}^{i_{2} + i_{3}}
p_{k_{3}}^{i_{4}}p_{k_{4}}^{i_{5}} +
p_{k_{1}}^{i_{1} + 
i_{2}}p_{k_{2}}^{i_{3}}
p_{k_{3}}^{i_{4}}p_{k_{4}}^{i_{5}}) \nonumber \\
& & + \sum_{k_{1} < k_{2} < k_{3}} \frac{1}{2^2} (
p_{k_{1}}^{i_{1}}p_{k_{2}}^{i_{2} + i_{3}}
p_{k_{3}}^{i_{4} + i_{5}} +
p_{k_{1}}^{i_{1} + i_{2}}p_{k_{2}}^{i_{3}}
p_{k_{3}}^{i_{4} + i_{5}} + 
p_{k_{1}}^{i_{1} + i_{2}}p_{k_{2}}^{i_{3} + i_{4}}
p_{k_{3}}^{i_{5}} ) \nonumber \\
& & - \sum_{k_{1} < k_{2}} \frac{1}{2^3} (
p_{k_{1}}^{i_{1}}p_{k_{2}}^{i_{2} + i_{3} + i_{4} + i_{5}} +
p_{k_{1}}^{i_{1} + i_{2} + i_{3} + i_{4}}p_{k_{2}}^{i_{5}} ) .
\end{eqnarray}
\end{example}

\eqreset
\section{Summary and discussion}

The present paper has reviewed the general scheme of higher-order decomposition
with special emphasis on determining equations of decomposition parameters.
The minimal independent determining equations (or sufficient conditions) 
are obtained by the requirement
that all the coefficients of the operator products $R_{i_{1}}R_{i_{2}}\cdots $
 ( in Suzuki's general scheme (4.8) ) corresponding to Lyndon words should vanish.
Here we regard $R_{j}$ as a word of length $j$.
Some explicit equations have been presented up to the 15 th order.
There are many related papers$^{26-44}$.

There remains a problem for solving numerically the determining equations 
for decomposition, for example, of the 10 th order$^{9}$, 
because of the reason mentioned in Appendix C  .

We conjecture the following algebraic relation will hold :
\begin{equation}
A_{m} / I_{m} \cong L^{( m )}( X ) 
\end{equation}
where \\
$A_{m}$ : $m$ th order sub-module generated by $m$-th order coefficients
$\{ g( \{ i_{j} \} ) \}$ .\\
$I_{m}$ : $m$ th order sub-module of the ideal generated by the 
coefficient $\{ g( \{ i_{j} \} ) \}$ of order lower than $m$.\\
$L^{( m )}(X)$ : $m$-th order sub-module of free Lie algebra 
generated by $X$ = $\{ R_{1}, R_{2}, R_{3}, \dots \}$
where we look upon $ R_{j} $ as a $j$ th order element.

Here we assume that the coefficients of Lyndon words are independent.
If this assumption dose not hold, (10.1) reduces to the following
relation :
 \begin{equation}
A_{m} / I_{m} \cong S^{( m )}( X ) 
\end{equation}
where $S^{( m )}( X )$ is a subset of $L^{( m )}(X)$ .
\\\\
%%%%%%%%%%%%%%%
{\bf Acknowledgements}

The present authors would like to thank Dr. R.I.Mclachlan for sending
them his preprints and critical comments.
One of the present authors ( Z. T. ) would like to thank the members
 of Suzuki research group, especially Dr. N. Hatano, for their kind
 advice on computers.
\\\\
%%%%%%%%%%%%%%
{\bf Note Added}\\
Recently Dr. E. Forest kindly sent to one of the authors (M.S.) a copy of 
the English translation of P.-V. Koseleff's Ph. D. thesis, in which a similar 
problem of decomposition has been studied using Lyndon words. 

In connection with Appendix C, we have recently generalized the functions of
1-variable, i.e. $a_{\alpha k}:=a_{k}[\alpha]$, 
$b_{\beta k}:=b_{k}[\beta]$, to the functions of many (two) variables
$a_{k}[i_{1},i_{2}]$ and $b_{k}[i_{1},i_{2}]$. 
Using this expression, the following relation holds:
\begin{eqnarray*}
\sum_{k} a_{k}[1]^2 p_{k}^{3} b_{k}[1,3] \equiv 2!f(1,1,3,1,3) \bmod D_{9}.
\end{eqnarray*}
Therefore, we can obtain desired determining equations for 9th order symmetric decomposition
by adding the equation $\sum_{k} a_{k}[1]^2 p_{k}^{3} b_{k}[1,3]=0 $ 
instead of $\sum_{k} a_{3k} p_{k}^{3} b_{1k}^{3}=0$ 
to Suzuki's determining equations$^{2}$.

In more detail, we have to explain explicit formulas on
 $a_{k}[i_{1},i_{2}]$ and $b_{k}[i_{1},i_{2}]$. 
Our strategy is to construct the function 
$a_{k}[i_{1},i_{2},\dots,i_{n}]$ and $b_{k}[i_{1},i_{2},\dots,i_{n}]$ 
satisfying the relation
\begin{eqnarray*}
a_{k}[i_{1}]a_{k}[i_{2}]\cdots a_{k}[i_{n}]=
\sum_{\sigma \in S_{n}}a_{k}[i_{\sigma(1)},i_{\sigma(2)},\dots,i_{\sigma(n)}]
\end{eqnarray*}
and
\begin{eqnarray*}
b_{k}[i_{1}]b_{k}[i_{2}]\cdots b_{k}[i_{n}]=
\sum_{\sigma \in S_{n}}b_{k}[i_{\sigma(1)},i_{\sigma(2)},\dots,i_{\sigma(n)}]
\end{eqnarray*}
Then, we find
\begin{eqnarray*}
a_{k}[\alpha,\beta]=
 \sum_{i<j<k}p_{i}^{\alpha}p_{j}^{\beta} 
 + \frac{1}{2}\sum_{j<k}p_{j}^{\alpha}p_{k}^{\beta} 
  + \frac{1}{2}\sum_{j<k}p_{j}^{\alpha + \beta} 
  + \frac{1}{2^{3}}p_{k}^{\alpha + \beta} ,
\end{eqnarray*}
and 
\begin{eqnarray*}
b_{k}[\alpha,\beta]=
 \sum_{k<i<j}p_{i}^{\alpha}p_{j}^{\beta} 
 + \frac{1}{2}\sum_{k<j}p_{k}^{\alpha}p_{j}^{\beta} 
  + \frac{1}{2}\sum_{k<j}p_{j}^{\alpha + \beta} 
  + \frac{1}{2^{3}}p_{k}^{\alpha + \beta} ,
\end{eqnarray*}
Furthermore, the following relation
\begin{eqnarray*}
\sum_{k} a_{k}[i_{1},\dots,i_{\alpha}] p_{k}^{s}b_{k}[j_{1},\dots,j_{\beta}] 
 \equiv f(i_{1},\dots, i_{\alpha}, s, j_{1},\dots, j_{\beta}) 
 \bmod E_{m}
\end{eqnarray*}
has been confirmed for small non-negative integers $\alpha$ and $\beta$. 
Here $E_{m}$ denotes a $K$-module
generated by $\{f(q_{1},q_{2},\dots,q_{n})\}$ for
 $1 \le n < \alpha +1 + \beta $ and 
$q_{1}+q_{2}+\cdots q_{n}=i_{1}+i_{2}+\cdots +i_{\alpha}+s+
j_{1}+\cdots + j_{\beta}=m $. The above relation is useful in numerical
calculations of decomposition parameters.

\eqreset
\renewcommand{\theequation}{A.\arabic{equation}}
\section*{A. Generalized Mclachlan's formula}

Now we explain the formula concerning the dimensionality of the 
sub-space of the free Lie algebra, though some special cases of it were first
proved by R. I. Mclachlan$^{25}$.

We introduce the following notation : 
\begin{equation}
M( \{ n_{j} \} ) := M( n_{1}, n_{2}, n_{3},\dots )
\end{equation} 
where r. h. s. of (A.1) is given by Witt's second formula$^{18}$ (3.5).
\begin{theorem}
 The number of independent $m$ th order commutators which are composed of 
$k_{i}$ th order {\rm (} $i = 1, 2, \dots, r$ {\rm )} independent commutators is
\begin{eqnarray}
\sum_{\sum_{i = 1}^{r} k_{i}n_{i} = m} M( \{ n_{i} \} ) &=&
 \sum_{\sum_{i = q + 1}^{r} k_{i}n_{i} = m} M( \{ n_{i} \} ) \nonumber \\
& & + \sum_{ \{ n_{l}^{( i ; t_{1}, t_{2}, \dots, t_{l - 1} )} \} \in {\it N_{m}}}
M( \{ n_{l}^{( i ; t_{1}, t_{2}, \dots, t_{l - 1} )} \} )
\end{eqnarray}
where
\begin{eqnarray}
{\it N_{m}} := \{ \{ n_{l}^{( i ; t_{1}, t_{2}, \dots, t_{l - 1} )} \} 
\in {\bf N^{\Lambda}} | \sum_{l = 1} \sum_{\{ t_{j} \}} \sum_{i = 1}^{q}
( k_{i} + \sum_{j = 1}^{l - 1} k_{t_{j}} ) 
n_{l}^{( i ; t_{1}, t_{2}, \dots, t_{l - 1} )} = m \} ; \nonumber \\
\Lambda = \{ ( l, ( i ; t_{1}, t_{2}, \dots, t_{l - 1} ) ) \} ; 
t_{j} \in \{ q + 1, q + 2, \dots, r \} ; 0 \le q \le r \}.
\end{eqnarray}
\end{theorem}
Proof. \\
Let $X = \{ x_{1}, x_{2}, \dots, x_{r} \}$, $S$ = 
$\{ x_{q + 1}, x_{q + 2}, \dots, x_{r} \}$ where $\{ x_{j} \}$ are 
indeterminates. The following relation holds$^{45, Chapter2, proposition10}$.\\
$L( X ) \cong L( S ) \oplus$ \{ free Lie algebra generated by ( $ad$ $(x_{t_{1}})$ $\cdot$ $ad$ $(x_{t_{2}})$ $\cdots$ $ad$ $(x_{t_{l-1}})$ )( $x_{i}$ ) | $l \ge 1$ ;  $x_{t_{1}}$,  $x_{t_{2}}$,$\dots$, $x_{t_{l-1}}$ $\in S$, $q+1 \le t_{j} \le r$ ;$x_{i}$ $\in$ $X-S$, $1 \le i \le q$ \} ; $(ad$ $(x_{t_{0}}))$ $(x_{i})$ = $x_{i}$.

If we look upon $x_{j}$ $(1 \le j \le r)$ as a $k_{j}$ th order basic Lie 
element, then  ( $ad$ $(x_{t_{1}})$ $\cdot$ $ad$ $(x_{t_{2}})$ $\cdots$ $ad$
$(x_{t_{l-1}})$ )( $x_{i}$ ) becomes $\sum_{j=1}^{l-1} k_{t_{j}}$ + $k_{i}$ th 
order basic Lie element. Therefore counting the dimensions of an $m$ th order
sub-module in both sides, we obtain the formula (A.2).

Theorem 11 contains the following formulas.
\begin{corollary}[M. Suzuki$^{19}$]
\begin{equation}
\sum_{\sum_{l} \{1+(l-1)k \}n_{l}=m} M(n_{1},n_{2},\dots)
=\sum_{j=1}^{[\frac{m}{k}]}
M(m-kj,j).
\end{equation}
\end{corollary}
Proof.\\
This is derived by putting $r=2$, $q=1$, $k_{1}=1$ and $k_{2}=k$ 
in Theorem 11.\\
This formulas is a generalization of Mclachlan's formulas, because 
\begin{equation}
 M_{2}(m)=\sum_{n_{1}+n_{2}=m} M(n_{1},n_{2})=\sum_{j=1}^{m} M(m-j,j), 
\end{equation}
namely we obtain the following formulas : 
\begin{corollary}[R. I. Mclachlan$^{25}$]
\begin{equation}
\sum_{n_{1}+2n_{2}+ \cdots mn_{m}=m} M(n_{1},n_{2},\dots,n_{m})
=\frac{1}{m} \sum_{d | m} \mu (d)2^{\frac{m}{d} }= M_{2}(m)
\end{equation}
with the M\"{o}bius function $\mu (d)$, using Witt's formula {\rm (3.4)}. 
\end{corollary}
Proof.\\
This is derived immediately by putting $k=1$ in (A.4).
\begin{corollary}[R. I. Mclachlan$^{25}$]
\begin{equation}
\sum_{\sum_{l=1} (2l-1)n_{l}=m} M(n_{1},n_{2},\dots)=\sum_{j=1}^{[\frac{m}{2}]}
M(m-2j,j).
\end{equation}
\end{corollary}
Proof.\\
This is an immediate consequence of (A.4) for $k=2$.

When $m=p$ ( prime ), we can simplify these formulas using the following 
theorem. 
\begin{theorem}
\begin{equation}
\sum_{\sum_{i=1}^{r} k_{i}n_{i}=p} M(n_{1},n_{2},\dots,n_{r})=
\sum_{\sum_{i=1}^{r} k_{i}n_{i}=p} \frac{1}{\sum_{j=1}^{r} n_{i}}\cdot 
\frac{(\sum_{i=1}^{r} n_{i})!}{n_{1}!n_{2}!\cdots n_{r}!}.
\end{equation}
\end{theorem}
Proof.\\
Let $n_{0}$ the greatest common measure of $\{n _{j} \}$.
Since $\sum_{i=1}^{r} n_{i}$ can be divided by $n_{0}$, 
the number $n_{0}$ have to
be 1 or $p$. If $n_{0}$ = $p$, then there exists $j$ $(1 \le j \le r)$
such that $n_{j}=p$, $k_{j}=1$ ; $n_{i}=0$, $i \ne j$. We may consider only 
the case $n_{0}$ = $1$, because $M(0,\dots,0,p,0,\dots,0)=0$ $(p \ge 2)$.
\begin{corollary}[R. I. Mclachlan$^{25}$]
\begin{equation}
\sum_{j=1}^{[\frac{p}{2}]} M(p-2j,j)=\sum_{j=1}^{[\frac{p}{2}]}
\frac{1}{p-j} \pmatrix{p-j \cr j}.
\end{equation}
\end{corollary}
Proof.\\
This is also easily derived by putting $r=2$, $q=1$, $k_{1}=1$ and $k_{2}=2$
 in Theorem 11.

\eqreset
\renewcommand{\theequation}{B.\arabic{equation}}
\section*{B. Other representations of the coefficients}

Equation (8.6) is given by the expansion of the following expression 
\begin{eqnarray}
& & n_{1}!n_{2}!\cdots n_{m}!\sum :(p_{k_{1}}^{i_{1}}+\frac{1}{2!}p_{k_{1}}^{i_{1}+
i_{2}}+\cdots+\frac{1}{n!}p_{k_{1}}^{i_{1}+i_{2}+\cdots+i_{n}}) \nonumber \\
& & \times (1+p_{k_{2}}^{i_{2}}+\frac{1}{2!}p_{k_{2}}^{i_{2}+
i_{3}}+\cdots+\frac{1}{(n-1)!}p_{k_{2}}^{i_{2}+i_{3}+\cdots+i_{n}}) \nonumber \\& & \times (1+p_{k_{3}}^{i_{3}}+\frac{1}{2!}p_{k_{3}}^{i_{3}+
i_{4}}+\cdots+\frac{1}{(n-2)!}p_{k_{3}}^{i_{3}+i_{4}+\cdots+i_{n}})
\cdots \nonumber \\
& & \times \cdots
(1+p_{k_{n-1}}^{i_{n-1}}+\frac{1}{2!}p_{k_{n-1}}^{i_{n-1}+i_{n}})
(1+p_{k_{n}}^{i_{n}}):.
\end{eqnarray}
Here, the notation $\sum : \dots :$ denotes the following rule:

The summation over $\{ k_{j} \}$ is performed under the following conditions:
\\
( 1 ) $i<j$ and $ k_{i}<k_{j}$ \ ,

and
\\
( 2 ) Each exponents $i_{j}$ appears only once in each term.
Otherwise, there is no contribution to the sum.
\begin{example}.
We have
\begin{equation}
\sum :p_{k_{1}}^{i_{1}} \cdot p_{k_{2}}^{i_{2}} \cdot \frac{1}{(n-2)!}
p_{k_{3}}^{i_{3}+i_{4}+\cdots+i_{n}}: =\frac{1}{(n-2)!}
\sum_{k_{1}<k_{2}<k_{3}}p_{k_{1}}^{i_{1}}p_{k_{2}}^{i_{2}}
p_{k_{3}}^{i_{3}+i_{4}+\cdots+i_{n}}, 
\end{equation}
and we have
\begin{equation}
\sum :\frac{1}{2!}p_{k_{1}}^{i_{1}+i_{2}} \cdot p_{k_{2}}^{i_{2}} \cdot
p_{k_{3}}^{i_{3}} \cdot \frac{1}{(n-3)!}p_{k_{4}}^{i_{4}+i_{5}+\cdots
+i_{n}}: =0 ,
\end{equation}
when  $i_{2}$ appears twice. We have also
\begin{equation}
\sum :p_{k_{1}}^{i_{1}} \cdot 1 \cdot p_{k_{3}}^{i_{3}} \cdot 
p_{k_{4}}^{i_{4}} \cdot \frac{1}{(n-4)!}
p_{k_{5}}^{i_{5}+i_{6}+\cdots+i_{n}}: = 0 
\end{equation}
when $i_{2}$ dose not appear. Similarly we obtain
\begin{eqnarray}
\sum :\frac{1}{2!}p_{k_{1}}^{i_{1}+i_{2}} \cdot 1 \cdot \frac{1}{4!} 
p_{k_{3}}^{i_{3}+i_{4}+i_{5}+i_{6}} \cdot 1 \cdot 1 \cdot 1 \cdot 
\frac{1}{(n-6)!}p_{k_{7}}^{i_{7}+i_{8}+\cdots+i_{n}}: \nonumber \\
= \frac{1}{2!4!(n-6)!}\sum_{k_{1}<k_{2}<k_{3}} p_{k_{1}}^{i_{1}+i_{2}}
p_{k_{3}}^{i_{3}+i_{4}+i_{5}+i_{6}}
p_{k_{7}}^{i_{7}+i_{8}+\cdots+i_{n}}. 
\end{eqnarray}
\end{example}

In a similar way, (8.15) is given by performing the above procedure in
 the following expression :
\begin{eqnarray}
\lefteqn{n_{1}!n_{2}!\cdots n_{m}!( -1 )^{M-n} \sum :((-1)^{(i_{1}-1)k_{1}}
p_{k_{1}}^{i_{1}}+\frac{1}{2!}(-1)^{(i_{1}+i_{2}-2)k_{1}}
p_{k_{1}}^{i_{1}+i_{2}}+\cdots} \nonumber \\ 
& & +\frac{1}{n!}(-1)^{(i_{1}+i_{2}+\cdots+i_{n}-n)k_{1}}
p_{k_{1}}^{i_{1}+i_{2}+\cdots+i_{n}}) 
(1+(-1)^{(i_{2}-1)k_{2}}p_{k_{2}}^{i_{2}}+ 
\frac{1}{2!}(-1)^{(i_{2}+i_{3}-2)k_{2}}p_{k_{2}}^{i_{2}+i_{3}} \nonumber \\
& & + \cdots+\frac{1}{(n-1)!}
(-1)^{(i_{2}+i_{3}+\cdots+i_{n}-(n-1))k_{2}}
p_{k_{2}}^{i_{2}+i_{3}+\cdots+i_{n}})
(1+(-1)^{(i_{3}-1)k_{3}}p_{k_{3}}^{i_{3}} \nonumber \\
& & + \frac{1}{2!}(-1)^{(i_{3}+i_{4}-2)k_{3}}p_{k_{3}}^{i_{3}+i_{4}}+
\frac{1}{(n-2)!}(-1)^{(i_{3}+i_{4}+\cdots+i_{n}-(n-2))k_{3}}
p_{k_{3}}^{i_{3}+i_{4}+\cdots+i_{n}})
\cdots \nonumber \\
& & \cdots
(1+(-1)^{(i_{n-1}-1)k_{n-1}}p_{k_{n-1}}^{i_{n-1}}+\frac{1}{2!}
(-1)^{(i_{n-2}+i_{n-1}-2)k_{n-1}}p_{k_{n-1}}^{i_{n-1}+i_{n}}) \nonumber \\
& & \times (1+(-1)^{(i_{n}-1)k_{n}}p_{k_{n}}^{i_{n}}): ,
\end{eqnarray}
where $M=i_{1}+i_{2}+\cdots+i_{n}$.

\eqreset
\renewcommand{\theequation}{C.\arabic{equation}}
\section*{C. Relation between Suzuki's representation and the present one}

The determining equations composed of $a_{\alpha k} = \sum_{j<k} p_{j}^{\alpha}+
\frac{1}{2}p_{k}^{\alpha}$ and $b_{\beta k} = \sum_{j>k} p_{j}^{\beta}+
\frac{1}{2}p_{k}^{\beta}$ which was proposed by M. Suzuki$^{2}$ can be 
expressed by our representations in a coset as follows.
\begin{example}
\ \\
$\circ$ the case $n=3$ $(\alpha+\beta+\gamma=m )$
\begin{equation}
\sum_{k} a_{\alpha k}a_{\beta k}a_{\gamma k} = 
\sum_{k} b_{\gamma k}b_{\beta k}b_{\alpha k} \equiv 0 \qquad \bmod C_{m},
\end{equation}
\begin{equation}
\sum_{k} a_{\alpha k}p_{k}^{\beta}b_{\gamma k} \equiv 
f(\alpha,\beta,\gamma) \qquad \bmod C_{m}.
\end{equation}
$\circ$ the case $n=5$ $(i_{1}+i_{2}+i_{3}+i_{4}+i_{5}=m)$ 
\begin{equation}
\sum_{k} a_{i_{1} k}a_{i_{2} k}a_{i_{3} k}a_{i_{4} k}a_{i_{5} k} =
\sum_{k} b_{i_{5} k}b_{i_{4} k}b_{i_{3} k}b_{i_{2} k}b_{i_{1} k}
 \equiv 0 \qquad \bmod D_{m}, 
\end{equation}
\begin{eqnarray}
\sum_{k} a_{i_{1} k}a_{i_{2} k}a_{i_{3} k}a_{i_{4} k}p_{k}^{i_{5}}
&=& \sum_{k} p_{k}^{i_{5}}b_{i_{4} k}b_{i_{3} k}b_{i_{2} k}b_{i_{1} k}
\nonumber \\ 
& \equiv & \sum_{\sigma \in S_{4}} 
f(i_{\sigma (1)},i_{\sigma (2)},i_{\sigma (3)},i_{\sigma (4)},i_{5})
\quad \bmod D_{m},
\end{eqnarray}
\begin{eqnarray}
\sum_{k} a_{i_{1} k}a_{i_{2} k}a_{i_{3} k}p_{k}^{i_{4}}b_{i_{5} k} 
&=& \sum_{k} a_{i_{5} k}p_{k}^{i_{4}}b_{i_{3} k}b_{i_{2} k}b_{i_{1} k} \nonumber \\
& \equiv &
\sum_{\sigma \in S_{3}} 
f(i_{\sigma (1)},i_{\sigma (2)},i_{\sigma (3)},i_{4},i_{5})
\qquad \bmod D_{m}, 
\end{eqnarray}
\begin{equation}
\sum_{k} a_{i_{1} k}a_{i_{2} k}p_{k}^{i_{3}}b_{i_{4} k}b_{i_{5} k} 
\equiv 
\sum_{\sigma \in X} 
f(i_{\sigma (1)},i_{\sigma (2)},i_{3},i_{\sigma (4)},i_{\sigma (5)}) \qquad
\bmod D_{m},
\end{equation}
with
\begin{equation}
 X=\{ \sigma \in S_{5} | \sigma( \{ 1,2\} )= \{1,2 \},\sigma(3)=3, \sigma(\{ 4,5 \} )=
\{ 4, 5 \} \}. 
\end{equation} 
Here, $C_{m}$ denotes a $K$-module generated by $f( m )$ and 
$D_{m}$ denotes a $K$-module generated both by $f( m )$ and by $f(\alpha,\beta,
\gamma)$ for 
$\alpha + \beta + \gamma = m$.
\end{example}
Suzuki's representation$^{2}$ up to the 8 th order has been confirmed to
 be correct.

The number of independent commutators composed of three $R_{1}$ and two
$R_{3}$ is $M(3,2)=2$. Thus independent coefficients are, for example,
$g(1,1,1,3,3)$ $=:$ $\beta _{1}$  and $g(1,1,3,1,3)$ $=:$ $\beta _{2}$.
In this case, if we examine every representation constructed by 
$\{ a_{\alpha j} \}$ and $\{ b_{\beta j} \}$ using Examples 6 and 12, we find 
that all determining equations can be expressed only by $\beta_{1}$
 under the condition in Theorem 7 for $m=9$  and for $ \bmod D_{9}$. In short, 
 Suzuki's determining equations for the 10 th
 order are missing $\beta _{2}$. So we can 
 not 
 obtain enough precision when we 
use the determining equations constructed only by $\{ a_{\alpha j} \}$ 
or $\{ b_{\beta j} \}$ in the case of higher order than the 9-th (or 
the 10-th order) on the symmetric decomposition. 
Therefore, we can obtain desired determining equations 
for 9 th order symmetric decomposition 
by adding the equation $g(1,1,3,1,3)=0$ to Suzuki's determining equations$^{2}$. \\
{\bf References}
\begin{enumerate}
\item M. Suzuki, {\it Phys. Lett.} {\bf A146 }(1990) 319.
\item M. Suzuki, {\it Phys. Lett.} {\bf A165 }(1992) 387. 
\item M. Suzuki, {\it J. Math. Phys. }{\bf 32 }(1991) 400.
\item H. Yoshida, {\it Phys. Lett.} {\bf A150} (1990) 262.
\item M. Suzuki, {\it J. Phys. Soc. Jpn.} {\bf 461} (1992) 3015.
\item M. Suzuki, {\it Physica} {\bf A191} (1992) 501.
\item M. Suzuki, {\it Proc. Japan Acad.} {\bf 69}, Ser.B, No.7, (1993) 161.
\item M. Suzuki, {\it Commun. Math. Phys.} {\bf 163} (1994) 491.
\item M. Suzuki, {\it Physica} {\bf A205} (1994) 65, and references cited therein.\item Z. Tsuboi, Master Thesis, Univ. of Tokyo (1995), in Japanese. 
\item H. F. Trotter, {\it Proc. Am. Math. Phys.} {\bf 10} (1959) 545.
\item M. Suzuki, {\it Commun. Math. Phys.} {\bf 51} (1976) 183.
\item M. Suzuki, {\it Prog. Theor. Phys.} {\bf 56} (1976) 1454.
\item M. Suzuki, S. Miyasita and A. Kuroda, {\it Prog. Theor. Phys.} {\bf 58}
      (1977) 1377.
\item M. Suzuki, {\it Phys. Lett.} {\bf A113} (1985) 299.
\item M. Suzuki, {\it J. Math. Phys. }{\bf 26} (1985) 601.
\item M. Suzuki, {\it J. Stat. Phys.} {\bf 26} (1986) 883.
\item W. Magnus, A.Karrass and D. Solitar, {\it Combinatorial group theory}
     (Dover, New York, 1976).
\item M. Suzuki, {\it Exponential Product Formula and Lie Algebra},in the 
     Proceedings Yamada Conference on 20 th Int. Colloq. on Group Theoretical
     Method in Physics, eds. A. Arima, T. Eguchi and N. Nakanishi ( World 
     Scientific, 1995 ).
\item R. Kubo, {\it J. Phys. Soc. Jpn.} {\bf 17} (1962) 1100.
\item R. D. Ruth, {\it IEEE Trans. Nucl. Sci.} {\bf NS-30} (1983) 2669; 
     F. Neri, Preprint (1988).
\item M. Suzuki and T. Yamauchi, {\it J. Math. Phys.} {\bf 34} (1993) 4892.
\item K. Aomoto, {\it On a unitary version of Suzuki's exponential product 
     formula J. Math. Soc. Japan} (in press).
\item M. Lothair, {\it Combinatorics on words} (Addison Wesly, 1983).
\item R. I. Mclachlan, {\it On the numerical integration of ordinary 
      differential equations by symmetric composition methods, SIAM J. Sci.
      Comp.} (1994). 
\item M. Suzuki, {\it Phys Lett.} {\bf A180} (1993) 232.
\item M. Suzuki, {\it Commun. Math. Phys.} {\bf 57} (1977) 193.
\item M. Suzuki, {\it Phys. Rev. }{\bf B31} (1985) 2957.
\item E. Forest and R. D. Ruth, {\it Physica} {\bf D43} (1990) 105.
\item E. Forest, {\it J. Math. Phys.} {\bf 31} (1990) 1133.
\item A. D. Bandrauk and H. Shen, {\it Chem. Phys. Lett.} {\bf 176} (1991) 428.
\item J. A. Oteo and J. Ros, {\it J. Phy. A Math. Gen.} {\bf 24} (1991) 5751.
\item W. Janke and T. Sauer, {\it Phys. Lett.} {\bf A165} (1992) 199.
\item N. Hatano and M. Suzuki, {\it Prog. Theor. Phys.} {\bf 85} (1991) 481.
\item M. Suzuki and K. Umeno, in {\it Computer Simulations in Condensed Matter 
     Phys.} {\bf 6}, eds. D. P. Landau, K. K. Mon and H. B. Shutter
     (Springer-Verlag, 1993) 74.
\item K. Umeno and M. Suzuki, {\it Phys. Lett.} {\bf A181} (1993) 383.
\item M. Glasner, D. Yevick and B. Hermansson, Mathl. {\it Comput. Modelling}
      {\bf 16} (1992) 177, and {\it Appl. Math. Lett}. 4 (1991) 85.
\item B. Hermansson and D. Yevick, {\it Opt. Lett.} {\bf 36} (1991) 354.
\item M. Glasner, D.Yevick and B. Hermansson, {\it Electronics Lett}. {\bf 27} 
      (1991) 475.
\item J. Candy and W. Rozmus, {\it J. Comp. Phys.} {\bf 92} (1991) 230.
\item Q. Sheng, {\it IMA J. Numer. Anal}. {\bf 9} (1989) 199, ibid. {\bf 14} 
      (1994) 27.
\item A. N. Drozdov, {\it Physica} {\bf A196} (1993) 283.
\item Z. Mei-Qing, {\it Phys. Lett.} {\bf A} (1993) 3.
\item H. Kobayasi, N. Hatano, M. Suzuki, {\it Physica} {\bf A211} (1994) 
      234-254.
\item N. Bourbaki, {\it \'{E}L\'{E}MENTS DE MATH\'{E}MATIQUE GROUPES ET 
      ALG\^^ {E}BRES DE LIE}
      ( HERMANN, Paris, 1972 )
\end{enumerate}
%\end{large}
\end{document}